\documentclass[floatfix,prd,twocolumn,letterpaper,lengthcheck,superscriptaddress,showpacs,amssymb,amsmath,amsfonts,aps,altaffilletter,nofootinbib,nopreprintnumbers,showpacs]{revtex4-1}
\pdfoutput=1
\usepackage{times}
\usepackage{mathptmx}
\usepackage{hyperref}
\usepackage{color}
\usepackage{graphicx}
\usepackage{float}
\usepackage[caption=false]{subfig}
\usepackage{amsmath}
\usepackage{multirow}
\usepackage{tabu}
\usepackage{tabularx}

\newcommand{\macro}[1]{#1}

\newcommand\rwRhoC{\macro{\ensuremath{\hat{\rho}_c}}}
\newcommand\fap{\ensuremath{\mathcal{F}}}
\newcommand{\dd}{\ensuremath{\mathrm{d}}}
\newcommand{\Msun}{\ensuremath{\mathrm{M}_\odot}}

\newcommand\OBSSTART{\macro{September 12}}
\newcommand\OBSEND{\macro{October 20}}
\newcommand\OBSYEAR{\macro{2015}}
\newcommand\OoneEND{\macro{January 12, 2016}}
\newcommand{\OBSDAYS}{\macro{\ensuremath{\macro{16}~\mathrm{days}}}}
\newcommand{\OBSHOURS}{\macro{\ensuremath{\macro{384}\,\mathrm{hr}}}}

\newcommand{\TheEvent}{GW150914}

\newcommand\OBSEVENTDATEMONTHDAYYEAR{\macro{September~14,~2015}}

\newcommand\OBSEVENTTIME{\macro{09:50:45}}
\newcommand\OBSEVENTTZ{\macro{UTC}}
\newcommand\OBSEVENTAPPROXCOMBINEDSNR{\macro{\ensuremath{24}}}
\newcommand\CBCEVENTIFAR{\macro{\ensuremath{203\,000}}}
\newcommand\PYCBCBCKLIVETIME{\macro{\ensuremath{608\,000}}}
\newcommand\CBCEVENTSIGMA{\macro{\ensuremath{5.1}}}
\newcommand\PYCBCHOWMANYTIMESLIDES{\macro{\ensuremath{\sim 10^7}}}
\newcommand\PYCBCTRIALFACTOR{\macro{3}}
\newcommand\CBCEVENTFAPBOUND{\macro{\ensuremath{< 2\times 10^{-7}}}}
\newcommand{\MONESCOMPACT}{\macro{\ensuremath{36_{-4}^{+5}}}}
\newcommand{\MTWOSCOMPACT}{\macro{\ensuremath{29_{-4}^{+4}}}}
\newcommand{\MCSCOMPACT}{\macro{\ensuremath{28_{-2}^{+2}}}}
\newcommand{\CHIEFFCOMPACT}{\macro{\ensuremath{-0.07_{-0.17}^{+0.16}}}}
\newcommand{\DISTANCECOMPACT}{\macro{\ensuremath{410_{-180}^{+160}}}}
\newcommand{\TIMEDELAYCOMPACT}{\macro{\ensuremath{6.9_{-0.4}^{+0.5}}}}
\newcommand{\SPINONELIMIT}{\macro{\ensuremath{0.7}}}
\newcommand{\REDSHIFTCOMPACT}{\macro{\ensuremath{0.09_{-0.04}^{+0.03}}}}
\newcommand\CBCEVENTFAR{\macro{\ensuremath{5 \times 10^{-6}}}}
\newcommand\PycbcEventNewSNR{\macro{\ensuremath{23.6}}}
\newcommand\CBCEventTemplateMassOne{\macro{\ensuremath{47.9}}}
\newcommand\CBCEventTemplateMassTwo{\macro{\ensuremath{36.6}}}
\newcommand\CBCEventPeakFrequency{\macro{\ensuremath{144}}}
\newcommand\CBCEventUTCTimeShort{\macro{14 September 2015 09:50:45}}
\newcommand\CBCEventTimeDiff{\macro{\ensuremath{7.1}}}
\newcommand\gstLALEventFAP{\macro{\ensuremath{1.4\times 10^{-11}}}}

\newcommand\SECONDMONDAY{\macro{LVT151012}}
\newcommand{\SecondTime}{October 12, 2015 at 09:54:43 UTC}
\newcommand\CBCSECONDEVENTIFAR{\macro{2.3}}
\newcommand\CBCSECONDEVENTFAP{\macro{0.02}}
\newcommand{\MONESCOMPACTSecondMonday}{\macro{\ensuremath{23_{-6}^{+18}}}}
\newcommand{\MTWOSCOMPACTSecondMonday}{\macro{\ensuremath{13_{-5}^{+4}}}}
\newcommand{\MCSCOMPACTSecondMonday}{\macro{\ensuremath{15_{-1}^{+1}}}}
\newcommand{\CHIEFFCOMPACTSecondMonday}{\macro{\ensuremath{0.0_{-0.2}^{+0.3}}}}
\newcommand{\DISTANCECOMPACTSecondMonday}{\macro{\ensuremath{1100_{-500}^{+500}}}}
\newcommand\CBCSECONDEVENTFAR{\macro{\ensuremath{0.44}}}
\newcommand\PyCBCSecondEventSigma{\macro{\ensuremath{2.1}}}
\newcommand\PyCBCSecondEventRhoC{\macro{\ensuremath{9.6}}}
\newcommand\gstLALSecondEventInverseFAR{\macro{\ensuremath{1.1}}}
\newcommand\gstLALSecondEventFAP{\macro{\ensuremath{0.05}}}
\newcommand\CBCSecondEventUTCTimeShort{\macro{12 October 2015 09:54:43}}

\newcommand\TotalAvailableCoincTime{\macro{18.4}}
\newcommand\TotalCoincAfterCATOne{\macro{17.5}}
\newcommand\CatOneVetoTime{\macro{20.7}}
\newcommand\CatTwoVetoTime{\macro{2}}

\newcommand{\pycbc}{PyCBC}

\newcommand{\gstlal}{GstLAL}
\newcommand{\gstlalXsq}{\ensuremath{\xi^2}}
\newcommand{\gstlalLR}{\ensuremath{\mathcal{L}}}
\newcommand{\gstlaltime}{17}
\newcommand{\signal}{\ensuremath{\mathrm{h}}}
\newcommand{\noise}{\ensuremath{\mathrm{n}}}
\newcommand\stats{\ensuremath{\mathbf{x}}}
\newcommand\Hone{\ensuremath{\mathrm{H}}}
\newcommand\Lone{\ensuremath{\mathrm{L}}}
\newcommand\intRegion{\ensuremath{S}}

\newcommand\FCHanford{\macro{341}}
\newcommand\FCLivingston{\macro{388}}
\newcommand\paramvary{\macro{10}}
\newcommand\imkappaTvary{\macro{0.1}}
\newcommand\mo{\macro{47.9}}
\newcommand\mt{\macro{36.6}}
\newcommand\chio{\macro{0.962}} 
\newcommand\chit{\macro{-0.900}}
\newcommand\imkappaTsnrlossH{\macro{2}}
\newcommand\imkappaTsnrlossL{\macro{5}}
\newcommand\seglen{\macro{2048}}

\begin{document}

\title[]{GW150914: First results from the search for binary black hole coalescence with Advanced LIGO}

\author{%
B.~P.~Abbott,$^{1}$  
R.~Abbott,$^{1}$  
T.~D.~Abbott,$^{2}$  
M.~R.~Abernathy,$^{1}$  
F.~Acernese,$^{3,4}$
K.~Ackley,$^{5}$  
C.~Adams,$^{6}$  
T.~Adams,$^{7}$
P.~Addesso,$^{3}$  
R.~X.~Adhikari,$^{1}$
% ...
\\
\medskip
(LIGO Scientific Collaboration and Virgo Collaboration) 
\\
\medskip
{{}$^{\dag}$Deceased, May 2015. {}$^{\ddag}$Deceased, March 2015. }%
}\noaffiliation
\affiliation {LIGO, California Institute of Technology, Pasadena, CA 91125, USA }
\affiliation {Louisiana State University, Baton Rouge, LA 70803, USA }
\affiliation {Universit\`a di Salerno, Fisciano, I-84084 Salerno, Italy }
\affiliation {INFN, Sezione di Napoli, Complesso Universitario di Monte S.Angelo, I-80126 Napoli, Italy }
\affiliation {University of Florida, Gainesville, FL 32611, USA }
\affiliation {LIGO Livingston Observatory, Livingston, LA 70754, USA }
\affiliation {Laboratoire d'Annecy-le-Vieux de Physique des Particules (LAPP), Universit\'e Savoie Mont Blanc, CNRS/IN2P3, F-74941 Annecy-le-Vieux, France }
\affiliation {Albert-Einstein-Institut, Max-Planck-Institut f\"ur Gravi\-ta\-tions\-physik, D-30167 Hannover, Germany }
\affiliation {Nikhef, Science Park, 1098 XG Amsterdam, Netherlands }
\affiliation {LIGO, Massachusetts Institute of Technology, Cambridge, MA 02139, USA }
\affiliation {Instituto Nacional de Pesquisas Espaciais, 12227-010 S\~{a}o Jos\'{e} dos Campos, S\~{a}o Paulo, Brazil }
\affiliation {INFN, Gran Sasso Science Institute, I-67100 L'Aquila, Italy }
\affiliation {INFN, Sezione di Roma Tor Vergata, I-00133 Roma, Italy }
\affiliation {Inter-University Centre for Astronomy and Astrophysics, Pune 411007, India }
\affiliation {International Centre for Theoretical Sciences, Tata Institute of Fundamental Research, Bangalore 560012, India }
\affiliation {University of Wisconsin-Milwaukee, Milwaukee, WI 53201, USA }
\affiliation {Leibniz Universit\"at Hannover, D-30167 Hannover, Germany }
\affiliation {Universit\`a di Pisa, I-56127 Pisa, Italy }
\affiliation {INFN, Sezione di Pisa, I-56127 Pisa, Italy }
\affiliation {Australian National University, Canberra, Australian Capital Territory 0200, Australia }
\affiliation {The University of Mississippi, University, MS 38677, USA }
\affiliation {California State University Fullerton, Fullerton, CA 92831, USA }
\affiliation {LAL, Universit\'e Paris-Sud, CNRS/IN2P3, Universit\'e Paris-Saclay, 91400 Orsay, France }
\affiliation {Chennai Mathematical Institute, Chennai 603103, India }
\affiliation {Universit\`a di Roma Tor Vergata, I-00133 Roma, Italy }
\affiliation {University of Southampton, Southampton SO17 1BJ, United Kingdom }
\affiliation {Universit\"at Hamburg, D-22761 Hamburg, Germany }
\affiliation {INFN, Sezione di Roma, I-00185 Roma, Italy }
\affiliation {Albert-Einstein-Institut, Max-Planck-Institut f\"ur Gravitations\-physik, D-14476 Potsdam-Golm, Germany }
\affiliation {APC, AstroParticule et Cosmologie, Universit\'e Paris Diderot, CNRS/IN2P3, CEA/Irfu, Observatoire de Paris, Sorbonne Paris Cit\'e, F-75205 Paris Cedex 13, France }
\affiliation {Montana State University, Bozeman, MT 59717, USA }
\affiliation {Universit\`a di Perugia, I-06123 Perugia, Italy }
\affiliation {INFN, Sezione di Perugia, I-06123 Perugia, Italy }
\affiliation {European Gravitational Observatory (EGO), I-56021 Cascina, Pisa, Italy }
\affiliation {Syracuse University, Syracuse, NY 13244, USA }
\affiliation {SUPA, University of Glasgow, Glasgow G12 8QQ, United Kingdom }
\affiliation {LIGO Hanford Observatory, Richland, WA 99352, USA }
\affiliation {Wigner RCP, RMKI, H-1121 Budapest, Konkoly Thege Mikl\'os \'ut 29-33, Hungary }
\affiliation {Columbia University, New York, NY 10027, USA }
\affiliation {Stanford University, Stanford, CA 94305, USA }
\affiliation {Universit\`a di Padova, Dipartimento di Fisica e Astronomia, I-35131 Padova, Italy }
\affiliation {INFN, Sezione di Padova, I-35131 Padova, Italy }
\affiliation {CAMK-PAN, 00-716 Warsaw, Poland }
\affiliation {University of Birmingham, Birmingham B15 2TT, United Kingdom }
\affiliation {Universit\`a degli Studi di Genova, I-16146 Genova, Italy }
\affiliation {INFN, Sezione di Genova, I-16146 Genova, Italy }
\affiliation {RRCAT, Indore MP 452013, India }
\affiliation {Faculty of Physics, Lomonosov Moscow State University, Moscow 119991, Russia }
\affiliation {SUPA, University of the West of Scotland, Paisley PA1 2BE, United Kingdom }
\affiliation {University of Western Australia, Crawley, Western Australia 6009, Australia }
\affiliation {Department of Astrophysics/IMAPP, Radboud University Nijmegen, 6500 GL Nijmegen, Netherlands }
\affiliation {Artemis, Universit\'e C\^ote d'Azur, CNRS, Observatoire C\^ote d'Azur, CS 34229, Nice cedex 4, France }
\affiliation {MTA E\"otv\"os University, ``Lendulet'' Astrophysics Research Group, Budapest 1117, Hungary }
\affiliation {Institut de Physique de Rennes, CNRS, Universit\'e de Rennes 1, F-35042 Rennes, France }
\affiliation {Washington State University, Pullman, WA 99164, USA }
\affiliation {Universit\`a degli Studi di Urbino ``Carlo Bo,'' I-61029 Urbino, Italy }
\affiliation {INFN, Sezione di Firenze, I-50019 Sesto Fiorentino, Firenze, Italy }
\affiliation {University of Oregon, Eugene, OR 97403, USA }
\affiliation {Laboratoire Kastler Brossel, UPMC-Sorbonne Universit\'es, CNRS, ENS-PSL Research University, Coll\`ege de France, F-75005 Paris, France }
\affiliation {Astronomical Observatory Warsaw University, 00-478 Warsaw, Poland }
\affiliation {VU University Amsterdam, 1081 HV Amsterdam, Netherlands }
\affiliation {University of Maryland, College Park, MD 20742, USA }
\affiliation {Center for Relativistic Astrophysics and School of Physics, Georgia Institute of Technology, Atlanta, GA 30332, USA }
\affiliation {Institut Lumi\`{e}re Mati\`{e}re, Universit\'{e} de Lyon, Universit\'{e} Claude Bernard Lyon 1, UMR CNRS 5306, 69622 Villeurbanne, France }
\affiliation {Laboratoire des Mat\'eriaux Avanc\'es (LMA), IN2P3/CNRS, Universit\'e de Lyon, F-69622 Villeurbanne, Lyon, France }
\affiliation {Universitat de les Illes Balears, IAC3---IEEC, E-07122 Palma de Mallorca, Spain }
\affiliation {Universit\`a di Napoli ``Federico II,'' Complesso Universitario di Monte S.Angelo, I-80126 Napoli, Italy }
\affiliation {NASA/Goddard Space Flight Center, Greenbelt, MD 20771, USA }
\affiliation {Canadian Institute for Theoretical Astrophysics, University of Toronto, Toronto, Ontario M5S 3H8, Canada }
\affiliation {Tsinghua University, Beijing 100084, China }
\affiliation {Texas Tech University, Lubbock, TX 79409, USA }
\affiliation {The Pennsylvania State University, University Park, PA 16802, USA }
\affiliation {National Tsing Hua University, Hsinchu City, 30013 Taiwan, Republic of China }
\affiliation {Charles Sturt University, Wagga Wagga, New South Wales 2678, Australia }
\affiliation {University of Chicago, Chicago, IL 60637, USA }
\affiliation {Caltech CaRT, Pasadena, CA 91125, USA }
\affiliation {Korea Institute of Science and Technology Information, Daejeon 305-806, Korea }
\affiliation {Carleton College, Northfield, MN 55057, USA }
\affiliation {Universit\`a di Roma ``La Sapienza,'' I-00185 Roma, Italy }
\affiliation {University of Brussels, Brussels 1050, Belgium }
\affiliation {Sonoma State University, Rohnert Park, CA 94928, USA }
\affiliation {Northwestern University, Evanston, IL 60208, USA }
\affiliation {University of Minnesota, Minneapolis, MN 55455, USA }
\affiliation {The University of Melbourne, Parkville, Victoria 3010, Australia }
\affiliation {The University of Texas Rio Grande Valley, Brownsville, TX 78520, USA }
\affiliation {The University of Sheffield, Sheffield S10 2TN, United Kingdom }
\affiliation {University of Sannio at Benevento, I-82100 Benevento, Italy and INFN, Sezione di Napoli, I-80100 Napoli, Italy }
\affiliation {Montclair State University, Montclair, NJ 07043, USA }
\affiliation {Universit\`a di Trento, Dipartimento di Fisica, I-38123 Povo, Trento, Italy }
\affiliation {INFN, Trento Institute for Fundamental Physics and Applications, I-38123 Povo, Trento, Italy }
\affiliation {Cardiff University, Cardiff CF24 3AA, United Kingdom }
\affiliation {National Astronomical Observatory of Japan, 2-21-1 Osawa, Mitaka, Tokyo 181-8588, Japan }
\affiliation {School of Mathematics, University of Edinburgh, Edinburgh EH9 3FD, United Kingdom }
\affiliation {Indian Institute of Technology, Gandhinagar Ahmedabad Gujarat 382424, India }
\affiliation {Institute for Plasma Research, Bhat, Gandhinagar 382428, India }
\affiliation {University of Szeged, D\'om t\'er 9, Szeged 6720, Hungary }
\affiliation {Embry-Riddle Aeronautical University, Prescott, AZ 86301, USA }
\affiliation {University of Michigan, Ann Arbor, MI 48109, USA }
\affiliation {Tata Institute of Fundamental Research, Mumbai 400005, India }
\affiliation {American University, Washington, D.C. 20016, USA }
\affiliation {University of Massachusetts-Amherst, Amherst, MA 01003, USA }
\affiliation {University of Adelaide, Adelaide, South Australia 5005, Australia }
\affiliation {West Virginia University, Morgantown, WV 26506, USA }
\affiliation {University of Bia{\l }ystok, 15-424 Bia{\l }ystok, Poland }
\affiliation {SUPA, University of Strathclyde, Glasgow G1 1XQ, United Kingdom }
\affiliation {IISER-TVM, CET Campus, Trivandrum Kerala 695016, India }
\affiliation {Institute of Applied Physics, Nizhny Novgorod, 603950, Russia }
\affiliation {Pusan National University, Busan 609-735, Korea }
\affiliation {Hanyang University, Seoul 133-791, Korea }
\affiliation {NCBJ, 05-400 \'Swierk-Otwock, Poland }
\affiliation {IM-PAN, 00-956 Warsaw, Poland }
\affiliation {Rochester Institute of Technology, Rochester, NY 14623, USA }
\affiliation {Monash University, Victoria 3800, Australia }
\affiliation {Seoul National University, Seoul 151-742, Korea }
\affiliation {University of Alabama in Huntsville, Huntsville, AL 35899, USA }
\affiliation {ESPCI, CNRS, F-75005 Paris, France }
\affiliation {Universit\`a di Camerino, Dipartimento di Fisica, I-62032 Camerino, Italy }
\affiliation {Southern University and A\&M College, Baton Rouge, LA 70813, USA }
\affiliation {College of William and Mary, Williamsburg, VA 23187, USA }
\affiliation {Instituto de F\'\i sica Te\'orica, University Estadual Paulista/ICTP South American Institute for Fundamental Research, S\~ao Paulo SP 01140-070, Brazil }
\affiliation {University of Cambridge, Cambridge CB2 1TN, United Kingdom }
\affiliation {IISER-Kolkata, Mohanpur, West Bengal 741252, India }
\affiliation {Rutherford Appleton Laboratory, HSIC, Chilton, Didcot, Oxon OX11 0QX, United Kingdom }
\affiliation {Whitman College, 345 Boyer Avenue, Walla Walla, WA 99362 USA }
\affiliation {National Institute for Mathematical Sciences, Daejeon 305-390, Korea }
\affiliation {Hobart and William Smith Colleges, Geneva, NY 14456, USA }
\affiliation {Janusz Gil Institute of Astronomy, University of Zielona G\'ora, 65-265 Zielona G\'ora, Poland }
\affiliation {Andrews University, Berrien Springs, MI 49104, USA }
\affiliation {Universit\`a di Siena, I-53100 Siena, Italy }
\affiliation {Trinity University, San Antonio, TX 78212, USA }
\affiliation {University of Washington, Seattle, WA 98195, USA }
\affiliation {Kenyon College, Gambier, OH 43022, USA }
\affiliation {Abilene Christian University, Abilene, TX 79699, USA }
\date{\today}

\begin{abstract}
On \OBSEVENTDATEMONTHDAYYEAR\ at \OBSEVENTTIME ~\OBSEVENTTZ\ the two detectors
of the Laser Interferometer Gravitational-wave Observatory (LIGO)
simultaneously observed the binary black hole merger \TheEvent{}.  We report the
results of a matched-filter search using relativistic models of compact-object binaries
that recovered \TheEvent{} as the most significant event during the
coincident observations between the two LIGO detectors from \OBSSTART\ to
\OBSEND,\,\OBSYEAR.  GW150914 was observed with a matched filter
signal-to-noise ratio of \OBSEVENTAPPROXCOMBINEDSNR\ and a false alarm rate
estimated to be less than 1 event per \CBCEVENTIFAR\ years, equivalent to a
significance greater than \CBCEVENTSIGMA\,$\sigma$. 
\end{abstract}

\maketitle

\section{Introduction}
\label{s:intro}
On \OBSEVENTDATEMONTHDAYYEAR\ at \OBSEVENTTIME ~\OBSEVENTTZ\ the LIGO Hanford,
WA, and Livingston, LA, observatories detected a signal from the binary black
hole merger \TheEvent{}~\cite{GW150914-DETECTION}.  The initial detection of
the event was made by low-latency searches for generic gravitational-wave
transients~\cite{GW150914-BURST}. We report the results of a matched-filter
search using relativistic models of compact binary coalescence waveforms that
recovered \TheEvent{} as the most significant event during the coincident
observations between the two LIGO detectors from \OBSSTART\ to
\OBSEND,\,\OBSYEAR. This is a subset of the data from Advanced LIGO's first
observational period that ended on \OoneEND. 

The binary coalescence search targets gravitational-wave emission from
compact-object binaries with individual masses from 1$\,$\Msun\, to
99$\,$\Msun, total mass less than 100$\,$\Msun\, and dimensionless spins up to
0.99. The search was performed using two independently implemented analyses,
referred to as \pycbc{}~\cite{Canton:2014ena,Usman:2015kfa,pycbc-github} and
\gstlal{}~\cite{Cannon:2011vi,Privitera:2013xza,gstlal-methods}.  These
analyses use a common set of template
waveforms~\cite{Taracchini:2013rva,Purrer:2015tud,Capano:2016uif}, but differ
in their implementations of matched
filtering~\cite{Allen:2005fk,Cannon:2010qh}, their use of detector
data-quality information~\cite{GW150914-DETCHAR}, the techniques used to
mitigate the effect of non-Gaussian noise transients in the
detector~\cite{Allen:2004gu,Cannon:2011vi}, and the methods for estimating the
noise background of the search~\cite{Usman:2015kfa,Cannon:2015gha}.  

\TheEvent{} was observed in both LIGO detectors~\cite{GW150914-DETECTORS}
with a time-of-arrival difference of 7\,ms, which is less than the 10\,ms inter-site propagation time, and a combined matched-filter
signal to noise ratio (SNR) of \OBSEVENTAPPROXCOMBINEDSNR.  The search
reported a false alarm rate estimated to be less than 1 event per
\CBCEVENTIFAR~years, equivalent to a significance greater than
\CBCEVENTSIGMA\,$\sigma$.  The basic features of the \TheEvent{} signal point
to it being produced by the coalescence of two black
holes~\cite{GW150914-DETECTION}.  The best-fit template parameters from the
search are consistent with detailed parameter estimation that identifies
\TheEvent{} as a near-equal mass black hole binary system with source-frame
masses {\MONESCOMPACT~\Msun} and {\MTWOSCOMPACT~\Msun} at the $90\%$ credible
level~\cite{GW150914-PARAMESTIM}.

The second most significant candidate event in the observation period
(referred to as \SECONDMONDAY{}) was reported on \SecondTime{} with a combined
matched-filter SNR of \PyCBCSecondEventRhoC.  The search reported a false
alarm rate of 1 per \CBCSECONDEVENTIFAR~years and a corresponding false alarm
probability of \CBCSECONDEVENTFAP{} for this candidate event.  Detector
characterization studies have not identified an instrumental or environmental
artifact as causing this candidate event~\cite{GW150914-DETCHAR}. However, its
false alarm probability is not sufficiently low to confidently claim this
candidate event as a signal~\cite{M1200055}.  Detailed waveform analysis of this candidate
event indicates that it is also a binary black hole merger with source frame
masses {\MONESCOMPACTSecondMonday~\Msun} and
{\MTWOSCOMPACTSecondMonday~\Msun}, if it is of astrophysical origin.
 
This paper is organized as follows: Sec.~\ref{s:overview} gives an overview of
the compact binary coalescence search and the methods used.
Sec.~\ref{s:pycbc} and Sec.~\ref{s:gstlal} describe the construction and
tuning of the two independently implemented analyses used in the search.
Sec.~\ref{s:results} presents the results of the search, and follow-up of the
two most significant candidate events, \TheEvent{} and \SECONDMONDAY{}.

\section{Search Description}
\label{s:overview}

The binary coalescence
search~\cite{thorne.k:1987,Sathyaprakash:1991mt,Cutler:1992tc,Finn:1992wt,Finn:1992xs,Dhurandhar:1992mw,Balasubramanian:1995bm,Flanagan:1997sx}
reported here targets gravitational waves from binary neutron stars, binary
black holes, and neutron star--black hole binaries, using matched
filtering~\cite{wainstein:1962} with waveforms predicted by general
relativity.  Both the \pycbc{} and \gstlal{} analyses correlate the detector
data with template waveforms that model the expected signal. The analyses
identify candidate events that are detected at both observatories consistent
with the $10$~ms inter-site propagation time. Events are assigned a
detection-statistic value that ranks their likelihood of being a
gravitational-wave signal. This detection statistic is compared to the
estimated detector noise background to determine the probability that a
candidate event is due to detector noise. 

We report on a search using coincident observations between the two Advanced
LIGO detectors~\cite{TheLIGOScientific:2014jea} in Hanford, WA (H1) and in
Livingston, LA (L1) from \OBSSTART\ to \OBSEND,\,\OBSYEAR. During these
$38.6$~days, the detectors were in coincident operation for a total of
\TotalAvailableCoincTime~days. Unstable instrumental operation and hardware
failures affected \CatOneVetoTime~hours of these coincident observations.
These data are discarded and the remaining \TotalCoincAfterCATOne~days are
used as input to the analyses~\cite{GW150914-DETCHAR}.  The analyses reduce
this time further by imposing a minimum length over which the detectors must
be operating stably; this is different between the two analysis (2064~s for
\pycbc{} and 512~s for \gstlal{}), as described
in Sec.~\ref{s:pycbc} and Sec.~\ref{s:gstlal}.  After applying this cut, the
\pycbc{} analysis searched $\OBSDAYS$ of coincident data and the \gstlal{}
analysis searched $\gstlaltime$~days of coincident data.  To prevent bias in
the results, the configuration and tuning of the analyses were determined
using data taken prior to \OBSSTART, \OBSYEAR.

A gravitational-wave signal incident on an interferometer alters its arm
lengths by $\delta L_x$ and $\delta L_y$, such that their measured difference
is $\Delta L(t) = \delta L_x - \delta L_y = h(t) L$, where $h(t)$ is the
gravitational-wave metric perturbation projected onto the detector, and $L$ is
the unperturbed arm length~\cite{Abramovici:1992ah}.  The strain is calibrated
by measuring the detector's response to test mass motion induced by photon
pressure from a modulated calibration laser beam~\cite{GW150914-CALIBRATION}.
Changes in the detector's thermal and alignment state cause small,
time-dependent systematic errors in the
calibration~\cite{GW150914-CALIBRATION}.  The calibration used for this search
does not include these time-dependent factors. Appendix \ref{s:calibration}
demonstrates that neglecting the time-dependent calibration factors does not
affect the result of this search.

The gravitational waveform $h(t)$ depends on the chirp mass of the binary,
$\mathcal{M} = (m_1 m_2)^{3/5} /
(m_1+m_2)^{1/5}$~\cite{PhysRev.131.435,Peters:1964}, the symmetric mass ratio
$\eta = (m_1 m_2) / (m_1 + m_2)^2$~\cite{Blanchet:1995ez}, and the angular
momentum of the compact objects $\chi_{1,2} = c{\bf
S}_{1,2}/Gm_{1,2}^2$~\cite{Kidder:1992fr,Kidder:1995zr} (the compact object's
dimensionless spin), where ${\bf S}_{1,2}$ is the angular momentum of the
compact objects. The effect of spin on the waveform depends also on the ratio
between the component objects' masses \cite{Blanchet:2013haa}.  Parameters
which affect the overall amplitude and phase of the signal as observed in the
detector are maximized over in the matched-filter search, but can be recovered
through full parameter estimation analysis~\cite{GW150914-PARAMESTIM}.  The
search parameter space is therefore defined by the limits placed on the compact
objects' masses and spins.  The minimum component masses of the search are
determined by the lowest expected neutron star mass, which we assume to be
$1\,\Msun$~\cite{Miller:2014aaa}. There is no known maximum black hole
mass~\cite{Belczynski:2014iua}, however we limit this search to binaries with a
total mass less than $M = m_1 + m_2 \le 100\,\Msun$. The LIGO detectors are
sensitive to higher mass binaries, however; the results of searches for
binaries that lie outside this search space will be reported in future
publications.
\begin{figure}[t]
\centering
\includegraphics[width=\columnwidth]{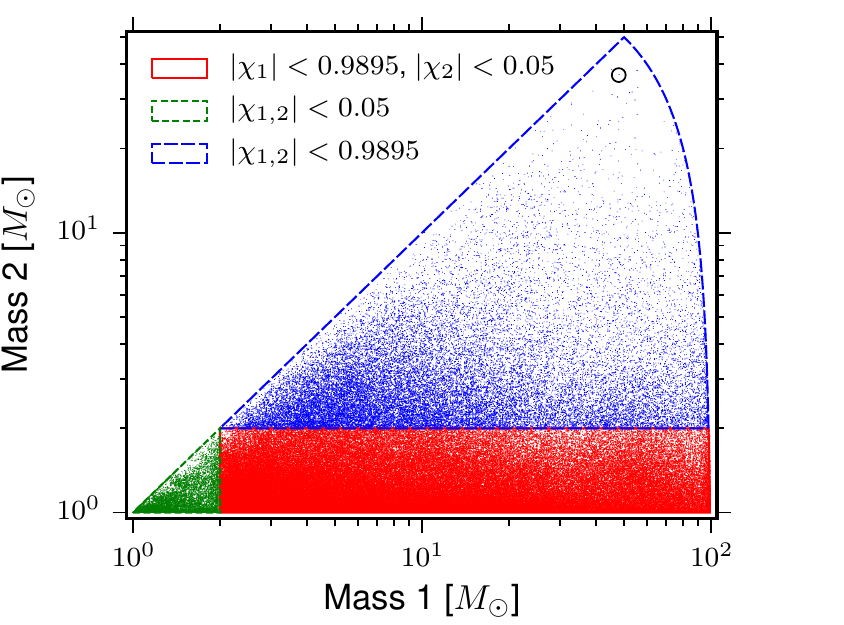}
\caption{The four-dimensional search parameter space covered by the template
bank shown projected into the component-mass plane, using the convention 
$m_1 > m_2$. The lines bound mass regions with different limits on the
dimensionless aligned-spin parameters $\chi_1$ and $\chi_2$. Each point
indicates the position of a template in the bank. The circle highlights the
template that best matches GW150914. This does not coincide with the best-fit
parameters due to the discrete nature of the template bank.
\label{fig:uberbank_boundaries}}
\end{figure}

The limit on the spins of the compact objects $\chi_{1,2}$ are informed by
radio and X-ray observations of compact-object binaries. The shortest observed
pulsar period in a double neutron star system is $22$~ms~\cite{Burgay:2003jj},
corresponding to a spin of $0.02$. Observations of X-ray binaries indicate
that astrophysical black holes may have near extremal
spins~\cite{McClintock:2013vwa}. In constructing the search, we assume that
compact objects with masses less than $2\,\Msun$ are neutron stars and we
limit the magnitude of the component object's spin to $0 \le \chi \le 0.05$.
For higher masses, the spin magnitude is limited to $0 \le \chi \le 0.9895$
with the upper limit set by our ability to generate valid template waveforms
at high spins~\cite{Taracchini:2013rva}. At current detector sensitivity,
limiting spins to $\chi_{1,2} \le 0.05$ for $m_{1,2} \le 2\, \Msun$ does not
reduce the search sensitivity for sources containing neutron stars with spins
up to $0.4$, the spin of the fastest-spinning millisecond
pulsar~\cite{Lorimer:2008se}.  Figure~\ref{fig:uberbank_boundaries} shows the
boundaries of the search parameter space in the component-mass plane, with the
boundaries on the mass-dependent spin limits indicated.

Since the parameters of signals are not known in advance, each detector's
output is filtered against a discrete bank of templates that span the search
target
space~\cite{Sathyaprakash:1991mt,Owen:1995tm,Owen:1998dk,Babak:2006ty,Cokelaer:2007kx}.
The placement of templates depends on the shape of the power spectrum of the
detector noise. Both analyses use a low-frequency cutoff of $30$~Hz for the search. 
The average noise power spectral density of the LIGO detectors
was measured over the period September 12 to September 26, 2015. The harmonic
mean of these noise spectra from the two detectors was used to place a single
template bank that was used for the duration of the
search~\cite{Keppel:2013uma,Usman:2015kfa}. The templates are placed using a combination of
geometric and stochastic
methods~\cite{Harry:2009ea,Brown:2012qf,Privitera:2013xza,Capano:2016uif}
such that the loss in matched-filter SNR caused by its discrete nature is
$\lesssim 3$\%.  Approximately 250,000 template waveforms are used to cover
this parameter space, as shown in Fig.~\ref{fig:uberbank_boundaries}. 

\begin{figure}[t]
\includegraphics[width=\columnwidth]{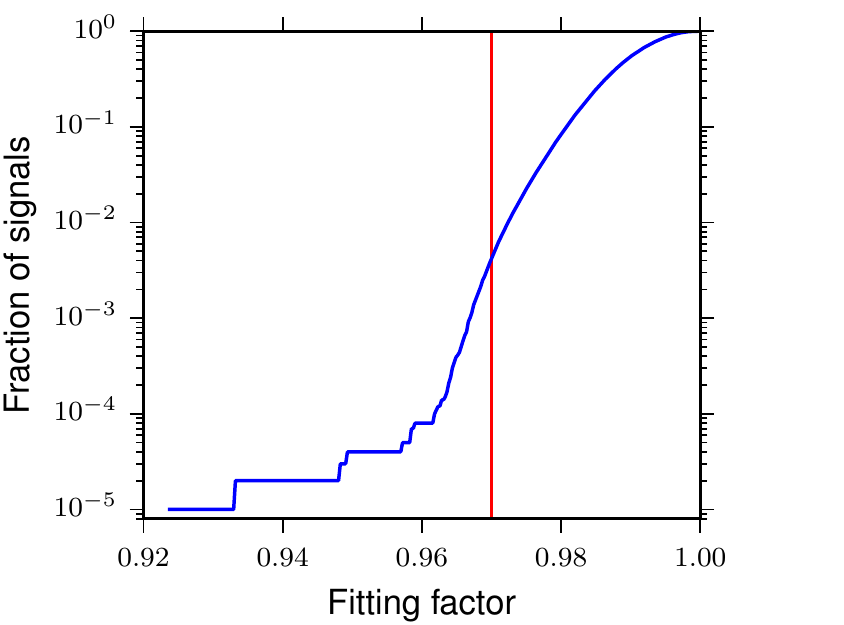} 
\caption{\label{fig:uberbank_effectualness}
Cumulative distribution of fitting factors obtained with the template bank for
a  population of simulated aligned-spin binary black hole signals.  Less than $1\%$ of the
signals have an matched-filter SNR loss greater than $3\%$, demonstrating that the
template bank has good coverage of the target search space.}
\end{figure}

The performance of the template bank is measured by the fitting
factor~\cite{Apostolatos:1996rf}; this is the fraction of the maximum
signal-to-noise ratio that can be recovered by the template bank for a signal
that lies within the region covered by the bank. The fitting factor is measured numerically
by simulating a signal and determining the maximum recovered matched-filter
SNR over the template bank. Figure~\ref{fig:uberbank_effectualness} shows the
resulting distribution of fitting factors obtained for the template bank over
the observation period. The loss in matched-filter SNR is less than $3\%$ for
more than $99$\% of the $10^5$ simulated signals.

\begin{figure}[t]
\includegraphics[width=\columnwidth]{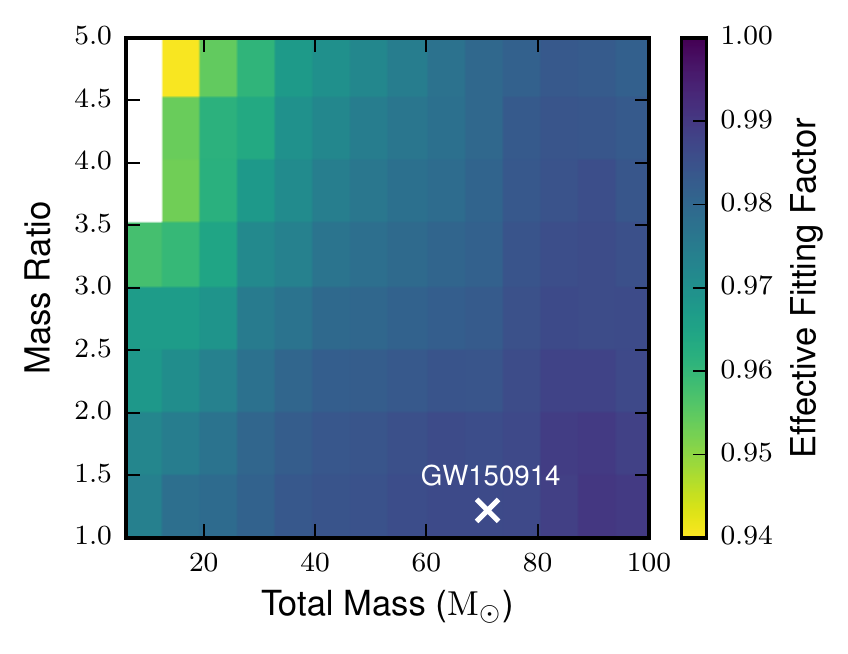}
\caption{\label{fig:precessing_effective_ff}
The effective fitting factor between simulated precessing binary black hole
signals and the template bank used for the search as a function of
detector-frame total mass and mass ratio, averaged over each rectangular tile.
The effective fitting factor gives the volume-averaged reduction in the
sensitive distance of the search at fixed matched-filter SNR due to mismatch
between the template bank and signals.  The cross shows the location of
\TheEvent{}. The high effective fitting factor near GW150914 demonstrates that
the aligned-spin template bank used in this search can effectively recover
systems with misaligned spins and similar masses to GW150914.}
\end{figure}

The template bank assumes that the spins of the two compact objects are
aligned with the orbital angular momentum. The resulting templates can
nonetheless effectively recover systems with misaligned spins in the
parameter-space region of GW150914.  To measure the effect of neglecting
precession in the template waveforms, we compute the effective fitting factor
which weights the fraction of the matched-filter SNR recovered by the
amplitude of the signal~\cite{BuonannoChenVallisneri:2003b}.  When a signal
with a poor orientation is projected onto the detector, the amplitude of the
signal may be too small to detect even if there was no mismatch between the
signal and the template; the weighting in the effective fitting accounts for
this.   Figure~\ref{fig:precessing_effective_ff} shows the effective fitting
factor for simulated signals from a population of simulated precessing binary
black holes that are uniform in co-moving
volume~\cite{Pan:2013rra,HarryetalInPrep}. The effective fitting factor is
lowest at high mass ratios and low total mass, where the effects of precession
are more pronounced. In the region close to the parameters of GW150914  the
aligned-spin template bank is sensitive to a large fraction of precessing
signals~\cite{HarryetalInPrep}.

In addition to possible gravitational-wave signals, the detector strain
contains a stationary noise background that primarily arises from photon shot
noise at high frequencies and seismic noise at low frequencies.  In the
mid-frequency range, detector commissioning has not yet reached the point
where test mass thermal noise dominates, and the noise at mid frequencies is
poorly
understood~\cite{GW150914-DETECTORS,GW150914-DETCHAR,InstrumentNoisePaper}.
The detector strain data also exhibits non-stationarity and non-Gaussian noise
transients that arise from a variety of instrumental or environmental
mechanisms. The measured strain $s(t)$ is the sum of possible
gravitational-wave signals $h(t)$ and the different types of detector noise
$n(t)$.

To monitor environmental disturbances and their influence on the detectors,
each observatory site is equipped with an array of
sensors~\cite{Effler:2014zpa}.  Auxiliary instrumental channels also record
the interferometer's operating point and the state of the detector's control
systems. Many noise transients have distinct signatures, visible in
environmental or auxiliary data channels that are not sensitive to
gravitational waves. When a noise source with known physical coupling between
these channels and the detector strain data is active, a data-quality veto is
created that is used to exclude these data from the
search~\cite{GW150914-DETCHAR}.  In the \gstlal{} analysis, time intervals
flagged by data quality vetoes are removed prior to the filtering. In the
\pycbc{} analysis, these data quality vetoes are applied after filtering.  A
total of \CatTwoVetoTime{}~hours is removed from the analysis by data quality
vetoes.  Despite these detector characterization investigations, the data
still contains non-stationary and non-Gaussian noise which can affect the
astrophysical sensitivity of the search. Both analyses implement methods to
identify loud, short-duration noise transients and remove them from the strain
data before filtering.

The \pycbc{} and \gstlal{} analyses calculate the matched-filter SNR for each
template and each detector's data~\cite{Allen:2005fk,Cannon2010}.  In the
\pycbc{} analysis, sources with total mass less than 4$\,$\Msun~are modeled by
computing the inspiral waveform accurate to third-and-a-half post-Newtonian
order~\cite{Blanchet:1995ez,Droz:1999qx,Blanchet:2004ek}.  To model systems
with total mass larger than 4$\,$\Msun,~we use templates based on the
effective-one-body (EOB) formalism~\cite{Buonanno:2000ef}, which combines
results from the Post-Newtonian
approach~\cite{Blanchet:1995ez,Blanchet:2004ek} with results from black hole
perturbation theory and numerical
relativity~\cite{Taracchini:2013rva,Puerrer:2014fza} to model the complete
inspiral, merger and ringdown waveform.  The waveform models used assume that
the spins of the merging objects are aligned with the orbital angular
momentum. The \gstlal{} analysis uses the same waveform families, but the
boundary between Post-Newtonian and EOB models is set at $\mathcal{M} = 1.74
\Msun$.  Both analyses identify maxima of the matched-filter SNR (triggers)
over the signal time of arrival.

To suppress large SNR values caused by non-Gaussian detector noise, the two
analyses calculate additional tests to quantify the agreement between the data
and the template.  The \pycbc{} analysis calculates a chi-squared statistic to
test whether the data in several different frequency bands are consistent with
the matching template~\cite{Allen:2004gu}.  The value of the chi-squared
statistic is used to compute a re-weighted SNR for each maxima.  The \gstlal{}
analysis computes a goodness-of-fit between the measured and expected SNR time
series for each trigger. The matched-filter SNR and goodness-of-fit values for
each trigger are used as parameters in the \gstlal{} ranking statistic.

Both analyses enforce coincidence between detectors by selecting trigger pairs
that occur within a $15\,$ms window and come from the same template.  The
$15\,$ms window is determined by the $10\,$ms inter-site propagation time plus
$5\,$ms for uncertainty in arrival time of weak signals.  The \pycbc{}
analyses discards any triggers that occur during the time of data-quality
vetoes prior to computing coincidence. The remaining coincident events are
ranked based on the quadrature sum of the re-weighted SNR from both
detectors~\cite{Usman:2015kfa}.  The \gstlal{} analysis ranks coincident
events using a likelihood ratio that quantifies the probability that a
particular set of concident trigger parameters is due to a signal versus the
probability of obtaining the same set of parameters from
noise~\cite{Cannon:2011vi}. 

The significance of a candidate event is determined by the search background.
This is the rate at which detector noise produces events with a
detection-statistic value equal to or higher than the candidate event (the
false alarm rate).  Estimating this background is challenging for two reasons:
the detector noise is non-stationary and non-Gaussian, so its properties must
be empirically determined; and it is not possible to shield the detector from
gravitational waves to directly measure a signal-free background.  The
specific procedure used to estimate the background is different for the two
analyses.

To measure the significance of candidate events, the \pycbc{} analysis
artificially shifts the time\-stamps of one detector's triggers by an offset
that is large compared to the inter-site propagation time, and a new set of
coincident events is produced based on this time-shifted data set. For
instrumental noise that is uncorrelated between detectors this is an effective
way to estimate the background.  To account for the search background noise
varying across the target signal space, candidate and background events are
divided into three search classes based on template length. To account for
having searched multiple classes, the measured significance is decreased by a
trials factor equal to the number of classes~\cite{Lyons:1900zz}.  

The \gstlal{} analysis measures the noise background using the distribution of
triggers that are not coincident in time. To account for the search background
noise varying across the target signal space, the analysis divides the
template bank into 248 bins.  Signals are assumed to be equally likely across
all bins and it is assumed that noise triggers are equally likely to produce a
given SNR and goodness-of-fit value  in any of the templates within a single
bin. The estimated probability density function for the likelihood statistic
is marginalized over the template bins and used to compute the probability of
obtaining a noise event with a likelihood value larger than that of a
candidate event.

The result of the independent analyses are two separate lists of candidate
events, with each candidate event assigned a false alarm probability and false
alarm rate. These quantities are used to determine if a gravitational-wave
signal is present in the search.  Simulated signals are added to the input
strain data to validate the analyses, as described in
Appendix~\ref{s:validation}.

\section{\pycbc{} Analysis}
\label{s:pycbc}

The \pycbc{} analysis~\cite{Canton:2014ena,Usman:2015kfa,pycbc-github}  uses
fundamentally the same
methods~\cite{Brown:2004pv,Allen:2005fk,Allen:2004gu,Brown:2005zs,Babak:2012zx,Brown:workflow,deelman2005pegasus,deelman2015pegasus,condor-practice,condor-dagman,scipy,numpy,matplotlib}
as those used to search for gravitational waves from compact binaries in the
initial LIGO and Virgo detector
era~\cite{Abbott:2003pj,Abbott:2005pe,Abbott:2005pf,Abbott:2005kq,Abbott:2007xi,Abbott:2007ai,Abbott:2009tt,Abbott:2009qj,Abadie:2010yba,Colaboration:2011np,Aasi:2012rja,Aasi:2014bqj},
with the improvements described in Refs.~\cite{Canton:2014ena,Usman:2015kfa}.
In this Section, we describe the configuration and tuning of the \pycbc{}
analysis used in this search.  To prevent bias in the search result, the
configuration of the analysis was determined using data taken prior to the
observation period searched. When GW150914 was discovered by the low-latency
transient searches~\cite{GW150914-DETECTION}, all tuning of the \pycbc{}
analysis was frozen to ensure that the reported false alarm probabilities are
unbiased. No information from the low-latency transient search is used in this
analysis.

Of the \TotalCoincAfterCATOne~days of data that are used as input to the
analysis, the \pycbc{} analysis discards times for which either of the LIGO
detectors is in their observation state for less than $2064$\,s; shorter
intervals are considered to be unstable detector operation by this analysis
and are removed from the observation time. After discarding time removed by
data-quality vetoes and periods when detector operation is considered unstable
the observation time remaining is $\OBSDAYS$.

For each template $h(t)$ and for the strain data from a single detector
$s(t)$, the analysis calculates the square of the matched-filter SNR defined
by~\cite{Allen:2005fk}
\begin{equation}
  \label{eqn:matched_filter_snr}
  \rho^2(t) \equiv \frac{1}{\langle h | h \rangle} \left| \langle s | h \rangle(t) \right|^2,
\end{equation}
where the correlation is defined by 
\begin{equation}
 \langle s|h\rangle(t) = 4 \int^\infty_0 \frac{\tilde{s}(f)\tilde{h}^*(f)}{S_n(f)} e^{2\pi i ft}\,\dd f\,,
 \label{eqn:matched_filter_innerprod}
\end{equation}
where $\tilde{s}(f)$ is the Fourier transform of the time domain quantity $s(t)$
given by
\begin{equation}
\tilde{s}(f) = \int_{-\infty}^{\infty} s(t) e^{-2\pi ift}\,\dd t.
\end{equation}
The quantity $S_n(|f|)$ is the one-sided average power spectral density of the
detector noise, which is re-calculated every 2048\,s (in contrast to the fixed
spectrum used in template bank construction).  Calculation of the
matched-filter SNR in the frequency domain allows the use of the
computationally efficient Fast Fourier Transform~\cite{intel-mkl,FFTW05}.  The
square of the matched-filter SNR in Eq.~(\ref{eqn:matched_filter_snr}) is
normalized by
\begin{equation}
 \langle h|h\rangle = 4 \int^\infty_0 \frac{\tilde{h}(f)\tilde{h}^*(f)}{S_n(f)} \,\dd f\,,
 \label{eqn:matched_filter_norm}
\end{equation}
so that its mean value is $2$, if $s(t)$ contains only stationary
noise~\cite{Cutler:1994}.

Non-Gaussian noise transients in the detector can produce extended periods of
elevated matched-filter SNR that increase the search
background~\cite{Usman:2015kfa}. To mitigate this, a time-frequency excess
power (burst) search~\cite{Robinet:2015om} is used to identify high-amplitude,
short-duration transients that are not flagged by data-quality vetoes. If the
burst search generates a trigger with a burst SNR exceeding $300$, the
\pycbc{} analysis vetoes these data by zeroing out $0.5$s of $s(t)$ centered
on the time of the trigger. The data is smoothly rolled off using a Tukey
window during the $0.25$\,s before and after the vetoed data. The threshold of
$300$ is chosen to be significantly higher than the burst SNR obtained from
plausible binary signals. For comparison, the burst SNR of GW150914 in the
excess power search is $\sim 10$. A total of $450$ burst-transient vetoes are
produced in the two detectors, resulting in $225$\,s of data removed from the
search. A time-frequency spectrogram of the data at the time of each
burst-transient veto was inspected to ensure that none of these windows
contained the signature of an extremely loud binary coalescence.
\begin{figure}[t]
 \vspace*{-0.15in}
 \hspace*{-0.02\columnwidth}
 \subfloat[H1, 16 $\chi^2$ bins]{%
 \includegraphics[height=0.29\textwidth]{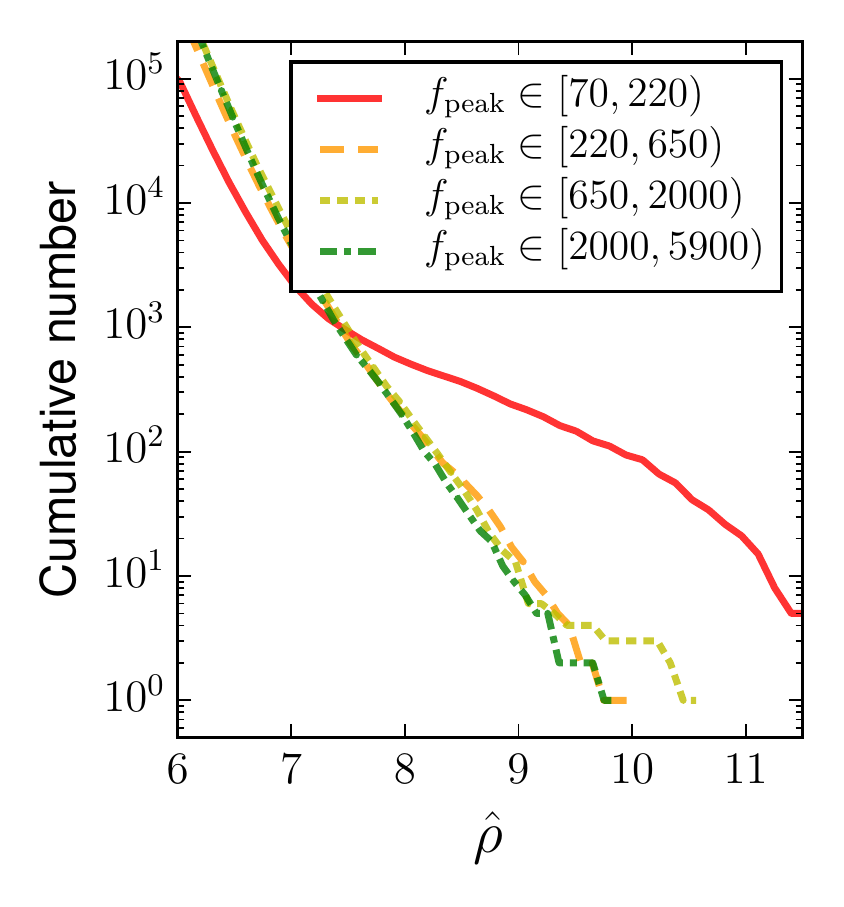}%
 }
 \subfloat[H1, optimized $\chi^2$ bins]{%
 \includegraphics[trim={0.65in 0 0 0}, clip, height=0.29\textwidth]{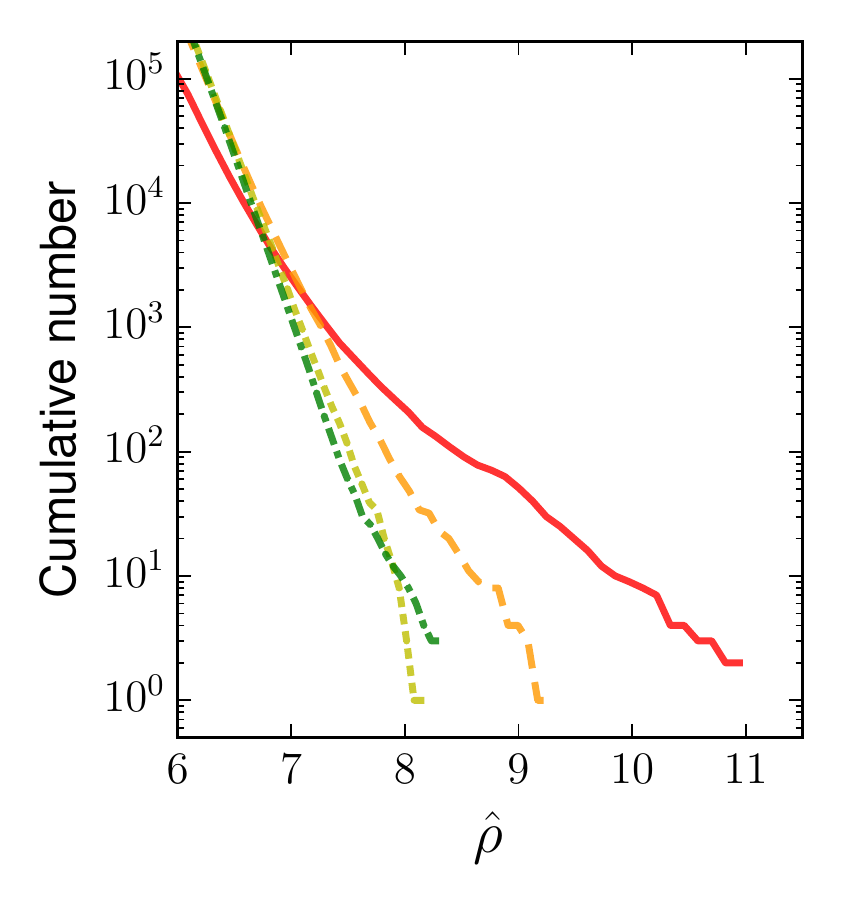}%
 }
\\
 \hspace*{-0.02\columnwidth}
 \subfloat[L1, 16 $\chi^2$ bins]{%
 \includegraphics[height=0.29\textwidth]{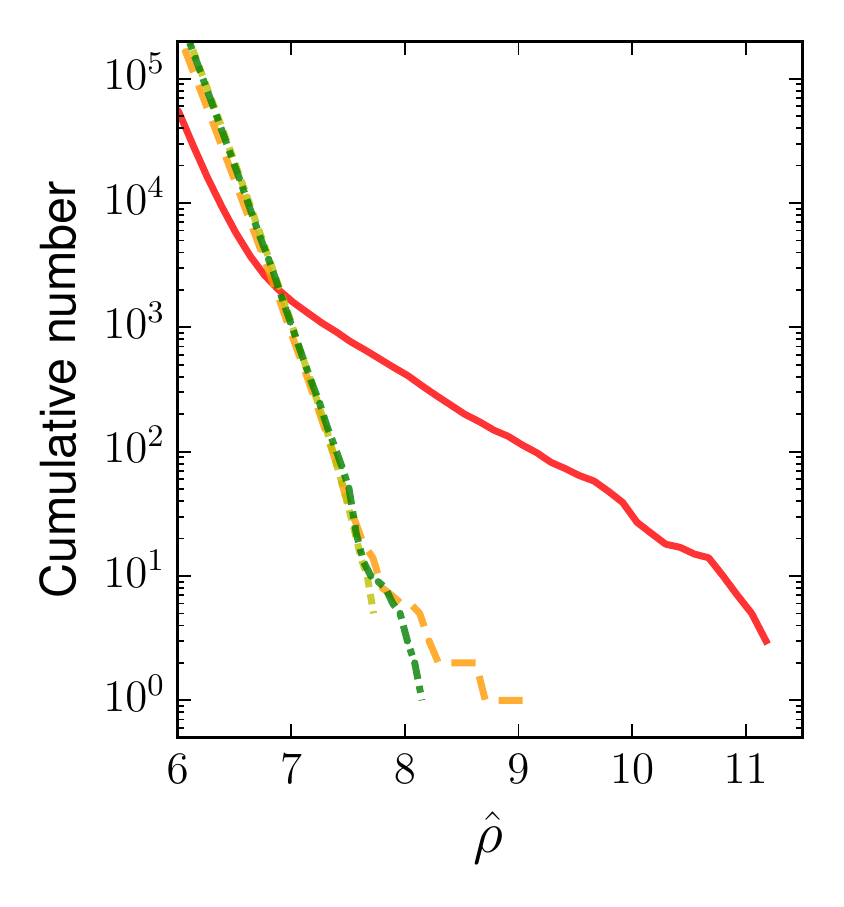}%
 }
 \subfloat[L1, optimized $\chi^2$ bins]{%
 \includegraphics[trim={0.65in 0 0 0}, clip, height=0.29\textwidth]{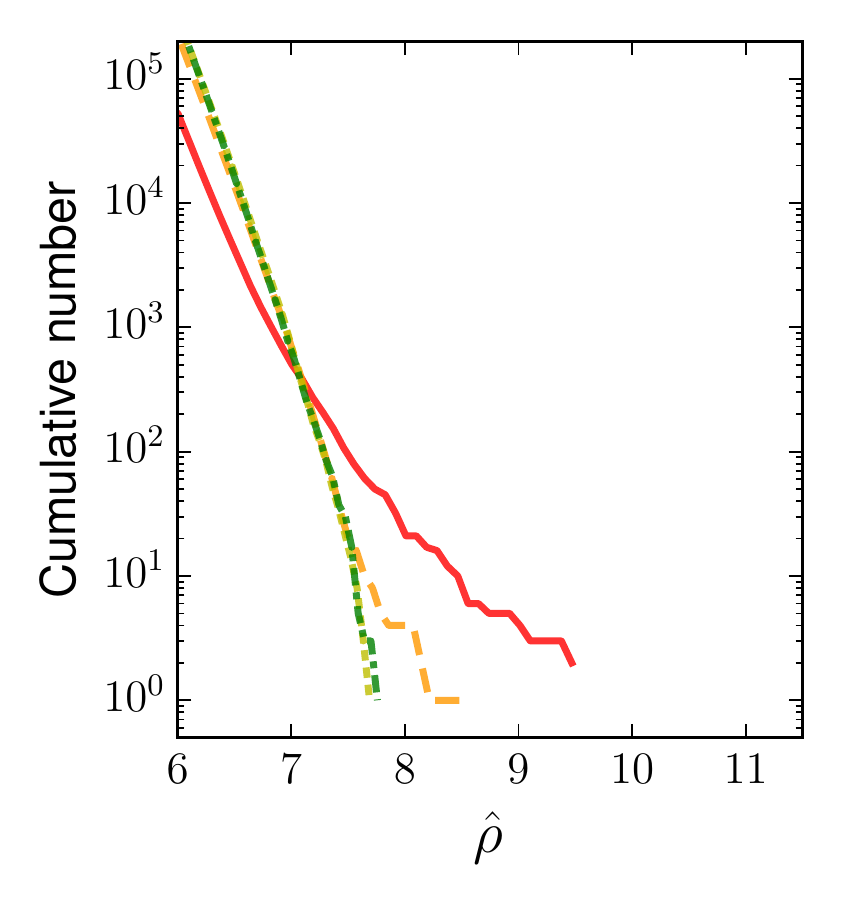}%
 }
\caption{\label{fig:chisq_and_template_bins}
Distributions of noise triggers over re-weighted SNR $\hat{\rho}$, for
Advanced LIGO engineering run data taken between September 2 and September 9,
2015.  Each line shows triggers from templates within a given range of
gravitational-wave frequency at maximum strain amplitude, $f_\text{peak}$.
Left: Triggers obtained from H1, L1 data respectively, using a fixed number of
$p=16$ frequency bands for the $\chi^2$ test.  Right: Triggers obtained with
the number of frequency bands determined by the function $p=\lfloor 0.4
(f_\mathrm{peak}/\mathrm{Hz})^{2/3} \rfloor$.  Note that while noise
distributions are suppressed over the whole template bank with the optimized
choice of $p$, the suppression is strongest for templates with lower
$f_\mathrm{peak}$ values. Templates that have a $f_\text{peak} < 220\,$Hz
produce a large tail of noise triggers with high re-weighted SNR even with the
improved $\chi^2$-squared test tuning, thus we separate these templates from
the rest of the bank when calculating the noise background.
}
\end{figure}

The analysis places a threshold of $5.5$ on the single-detector matched-filter
SNR and identifies maxima of $\rho(t)$  with respect to the time of arrival of
the signal. For each maximum we calculate a chi-squared statistic to determine
whether the data in several different frequency bands are consistent with the
matching template~\cite{Allen:2004gu}. Given a specific number of frequency
bands $p$, the value of the reduced $\chi_r^2$ is given by
\begin{equation}
\chi_r^2 = \frac{p}{2p-2} \frac{1}{\langle h | h \rangle} \sum_{i=1}^p  \left|\langle s | h_i\rangle - \frac{\langle s | h\rangle}{p}\right|^2,
\label{eqn:pycbc_chisq}
\end{equation}
where $h_i$ is the sub-template corresponding to the $i$-th
frequency band.  Values of $\chi_r^2$ near unity indicate that the signal is
consistent with a coalescence. To suppress triggers from noise transients with
large matched-filter SNR, $\rho(t)$ is re-weighted
by~\cite{Colaboration:2011np,Babak:2012zx}
\begin{equation}
\hat{\rho} = \left\{\begin{array}{lr}
\rho\left/\left[(1+(\chi^2_r)^3)/2\right]^\frac{1}{6}\right., & \text{if } \chi_r^2 > 1, \\
\rho, & \text{if } \chi_r^2 \le 1.
\end{array}\right.
\end{equation}
Triggers that have a re-weighted SNR $\hat{\rho} < 5$ or that occur during
times subject to data-quality vetoes are discarded.

The template waveforms span a wide region of time-frequency parameter space
and the susceptibility of the analysis to a particular type of noise transient
can vary across the search space. This is demonstrated in
Fig.~\ref{fig:chisq_and_template_bins} which shows the cumulative number of
noise triggers as a function of re-weighted SNR for Advanced LIGO engineering
run data taken between September 2 and September 9, 2015. The response of the
template bank to noise transients is well characterized by the
gravitational-wave frequency at the template's peak amplitude,
$f_\mathrm{peak}$. Waveforms with a lower peak frequency have less cycles in
the detector's most sensitive frequency band from
$30$--$2000$\,Hz~\cite{GW150914-DETECTORS,InstrumentNoisePaper}, and so are
less easily distinguished from noise transients by the re-weighted SNR. 

The number of bins in the $\chi^2$ test is a tunable parameter in the
analysis~\cite{Usman:2015kfa}. Previous searches used a fixed number of
bins~\cite{Babak:2005kv} with the most recent Initial LIGO and Virgo searches
using $p=16$ bins for all templates~\cite{Colaboration:2011np,Aasi:2012rja}.
Investigations on data from LIGO's sixth science
run~\cite{AlexNitzThesis,Aasi:2012rja} showed that better noise rejection is
achieved with a template-dependent number of bins.  The left two panels of
Fig.~\ref{fig:chisq_and_template_bins} show the cumulative number of noise
triggers with $p = 16$ bins used in the $\chi^2$ test.  Empirically, we find
that choosing the number of bins according to
\begin{equation}
p=\lfloor 0.4 (f_\mathrm{peak}/\mathrm{Hz})^{2/3}\rfloor
\label{eq:chisq_bins}
\end{equation}
gives better suppression of noise transients in Advanced LIGO data, as shown
in the right panels of Fig.~\ref{fig:chisq_and_template_bins}.

The \pycbc{} analysis enforces signal coincidence between detectors by
selecting trigger pairs that occur within a $15\,$ms window and come from the
same template.   We rank coincident events based on the quadrature sum
$\hat{\rho}_c$ of the $\hat{\rho}$ from both detectors~\cite{Usman:2015kfa}.
The final step of the analysis is to cluster the coincident events, by
selecting those with the largest value of $\hat{\rho}_c$ in each time window
of $10$\,s. Any other events in the same time window are discarded.  This
ensures that a loud signal or transient noise artifact gives rise to at most
one candidate event~\cite{Usman:2015kfa}. 

The significance of a candidate event is determined by the rate at which
detector noise produces events with a detection-statistic value equal to or
higher than that of the candidate event. To measure this, the analysis creates
a ``background data set" by artificially shifting the time\-stamps of one
detector's triggers by many multiples of $0.1$\,s and computing a new set of
coincident events.  Since the time offset used is always larger than the
time-coincidence window, coincident signals do not contribute to this
background. Under the assumption that noise is not correlated between the
detectors~\cite{GW150914-DETCHAR}, this method provides an unbiased estimate
of the noise background of the analysis. 

To account for the noise background varying across the target signal space,
candidate and background events are divided into different search classes
based on template length.  Based on empirical tuning using Advanced LIGO
engineering run data taken between September 2 and September 9, 2015, we
divide the template space into three classes according to: (i) $\mathcal{M} <
1.74\,\Msun$; (ii) $\mathcal{M}
\geq 1.74\,\Msun$ and $f_\mathrm{peak} \geq 220\,$Hz; (iii) $\mathcal{M} \geq 1.74\,\Msun$
and $f_\mathrm{peak} < 220\,$Hz.  The significance of candidate events is
measured against the background from the same class.  For each candidate
event, we compute the false alarm probability \fap{}. This is the probability
of finding one or more noise background events in the observation time with a
detection-statistic value above that of the candidate event, given
by~\cite{Usman:2015kfa, Capano:2016uif}
\begin{equation} 
\begin{split}
\label{eq:pycbc_slide_fap}
 \fap(\hat{\rho}_c) \equiv P(\geq 1\;\text{noise event above}\;\hat{\rho}_c|T, T_b) = \\
 1 - \exp\left[ -T \frac{1+n_b(\hat{\rho}_c)}{T_b} \right],
\end{split}
\end{equation}
where $T$ is the observation time of the search, $T_b$ is the background time,
and $n_b(\hat{\rho}_c)$ is the number of noise background triggers above the
candidate event's re-weighted SNR $\hat{\rho}_c$. 

\begin{figure}[!t]
\hspace*{-0.1in}
\includegraphics[width=\columnwidth]{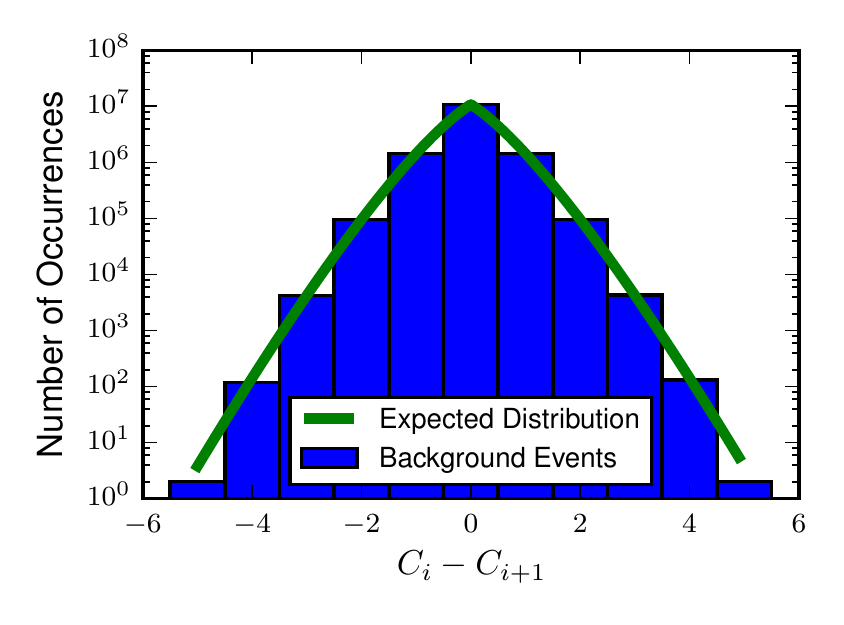}
\caption{\label{fig:slide_corr} 
The distribution of the differences in the number of events between 
consecutive time shifts, where $C_i$ denotes the number of events in the 
$i$th time shift.
The green line shows the predicted distribution for independent Poisson
processes with means equal to the average event rate per time shift. 
The blue histogram shows the distribution obtained from time-shifted 
analyses.  The variance of the time-shifted background distribution is 
1.996, consistent with the predicted variance of 2. The distribution 
of background event counts in adjacent time shifts is well modeled by 
independent Poisson processes.}
\end{figure}

Eq.~(\ref{eq:pycbc_slide_fap}) is derived assuming Poisson statistics for the
counts of time-shifted background events, and for the count of coincident
noise events in the search~\cite{Usman:2015kfa, Capano:2016uif}.  This
assumption requires that different time-shifted analyses (i.e.\ with different
relative shifts between detectors) give independent realizations of a counting
experiment for noise background events. We expect different time shifts to
yield independent event counts since the $0.1$\,s offset time is greater than
the $10$\,ms gravitational-wave travel time between the sites plus the $\sim
1$\,ms autocorrelation length of the templates. To test the independence of
event counts over different time shifts over this observation period, we
compute the differences in the number of background events having $\rwRhoC >
9$ between consecutive time shifts.  Figure~\ref{fig:slide_corr} shows that the
measured differences on these data follow the expected distribution for the
difference between two independent Poisson random
variables~\cite{Skellam:1946}, confirming the independence of time shifted
event counts. 

If a candidate event's detection-statistic value is larger than that of any
noise background event, as is the case for GW150914, then the \pycbc{}
analysis places an upper bound on the candidate's false alarm probability.
After discarding time removed by data-quality vetoes and periods when the
detector is in stable operation for less than $2064$~seconds, the total
observation time remaining is $T= \OBSDAYS$.  Repeating the time-shift
procedure {\PYCBCHOWMANYTIMESLIDES} times on these data produces a noise
background analysis time equivalent to $T_b=\PYCBCBCKLIVETIME$ years.  Thus,
the smallest false alarm probability that can be estimated in this analysis is
approximately $\fap = 7\times 10^{-8}$. Since we treat the search parameter
space as \PYCBCTRIALFACTOR{} independent classes, each of which may generate a
false positive result, this value should be multiplied by a trials factor or
look-elsewhere effect~\cite{Lyons:1900zz} of \PYCBCTRIALFACTOR{}, resulting in
a minimum measurable false alarm probability of $\fap = 2\times 10^{-7}$.  The
results of the \pycbc{} analysis are described in Sec.~\ref{s:results}.

\section{\gstlal{} Analysis}
\label{s:gstlal}

The \gstlal{}~\cite{GstLAL} analysis implements a time-domain matched filter
search~\cite{Cannon:2011vi} using techinques that were developed to perform
the near real-time compact-object binary
searches~\cite{Privitera:2013xza,gstlal-methods}. To accomplish this, the data
$s(t)$ and templates $h(t)$ are each whitened in the frequency domain by
dividing them by an estimate of the power spectral density of the detector
noise.  An estimate of the stationary noise amplitude spectrum is obtained
with a combined median--geometric-mean modification of Welch's
method~\cite{gstlal-methods}.  This procedure is applied piece-wise on
overlapping Hann-windowed time-domain blocks that are subsequently summed
together to yield a continuous whitened time series $s_w(t)$.  The time-domain
whitened template $h_w(t)$ is then convolved with the whitened data $s_w(t)$
to obtain the matched-filter SNR time series $\rho(t)$ for each template.  By
the convolution theorem, $\rho(t)$ obtained in this manner is the same as the
$\rho(t)$ obtained by frequency domain filtering in
Eq.~\eqref{eqn:matched_filter_snr}.

Of the \TotalCoincAfterCATOne~days of data that are used as input to the
analysis, the \gstlal{} analysis discards times for which either of the LIGO
detectors is in their observation state for less than $512$~s in duration.
Shorter intervals are considered to be unstable detector operation by this
analysis and are removed from the observation time. After discarding time
removed by data-quality vetoes and periods when the detector operation is
considered unstable the observation time remaining is $\gstlaltime{}$~days.
To remove loud, short-duration noise transients, any excursions in the
whitened data that are greater than $50\sigma$ are removed with 0.25~s
padding.  The intervals of $s_w(t)$ vetoed in this way are replaced with
zeros.  The cleaned whitened data is the input to the matched filtering stage.

Adjacent waveforms in the template bank are highly correlated.  The \gstlal{}
analysis takes advantage of this to reduce the computational cost of the
time-domain correlation. The templates are grouped by chirp mass and spin into
248 bins of $\sim1000$ templates each. Within each bin, a reduced set of
orthonormal basis functions $\hat{h}(t)$ is obtained via a singular value
decomposition of the whitened templates.  We find that the ratio of the number
of orthonormal basis functions to the number of input waveforms is $\sim$0.01
-- 0.10, indicating a significant redundancy in each bin.  The set of
$\hat{h}(t)$ in each bin is convolved with the whitened data; linear
combinations of the resulting time series are then used to reconstruct the
matched-filter SNR time series for each template. This decomposition allows
for computationally-efficient time-domain filtering and reproduces the
frequency-domain matched filter $\rho(t)$ to within 0.1\%~\cite{Cannon2010,
Cannon:2011vi, Cannon:2011xk}.

Peaks in the matched-filter SNR for each detector and each template are
identified over $1$~s windows. If the peak is above a matched-filter SNR of 4,
it is recorded as a trigger.  For each trigger, the matched-filter SNR time
series around the trigger is checked for consistency with a signal by
comparing the template's autocorrelation function $R(t)$ to the matched-filter
SNR time series $\rho(t)$. The residual found after subtracting the
autocorrelation function forms a goodness-of-fit test,
\begin{equation}
  \gstlalXsq = \frac{1}{\mu} \int_{t_p-\delta t}^{t_p+\delta t} \mathrm{d} t |\rho(t_p) R(t) - \rho(t)|^2,
\label{eq:autochisq_integrated}
\end{equation}
where $t_p$ is the time at the peak matched-filter SNR $\rho(t_p)$, and
$\delta t$ is a tunable parameter. A suitable value for $\delta t$ was found
to be 85.45\,ms (175 samples at a $2048\,$Hz sampling rate). The quantity
$\mu$ normalizes \gstlalXsq{}  such that a well-fit signal has a mean value of
$1$ in Gaussian noise~\cite{gstlal-methods}. The \gstlalXsq{} value is
recorded with the trigger.

Each trigger is checked for time coincidence with triggers from the same
template in the other detector.  If two triggers occur from the same template
within $15$~ms in both detectors, a coincident event is recorded.  Coincident
events are ranked according to a multidimensional likelihood ratio
\gstlalLR{}~\cite{0264-9381-25-10-105024, Cannon:2015gha}, then clustered in a
$\pm 4\,$s time window. The likelihood ratio ranks candidate events by the
ratio of the probability of observing matched-filter SNR and $\gstlalXsq$ from
signals (\signal{}) versus obtaining the same parameters from noise
(\noise{}).  Since the orthonormal filter decomposition already groups
templates into regions with high overlap, we expect templates in each group to
respond similarly to noise.  We use the template group $\theta_i$ as an
additional parameter in the likelihood ratio to account for how different
regions of the compact binary parameter space are affected differently by
noise processes.  The likelihood ratio is thus:
\begin{equation}
\gstlalLR = \frac{
    p(\stats_{\Hone}, \stats_{\Lone}, D_{\Hone}, D_{\Lone}|\theta_i, \signal)}{
    p(\stats_{\Hone}|\theta_i, \noise)p(\stats_{\Lone}|\theta_i, \noise)},
\label{eq:gstlalLR}
\end{equation}
where $\stats_d = \{\rho_d, \gstlalXsq_d\}$ are the matched-filter SNR and
\gstlalXsq{} in each detector, and $D$ is a parameter that measures the
distance sensitivity of the given detector during the time of a trigger. 

The numerator of the likelihood ratio is generated using an astrophysical
model of signals distributed isotropically in the nearby Universe to calculate
the joint SNR distribution in the two detectors~\cite{Cannon:2015gha}.  The
\gstlalXsq{} distribution for the signal hypothesis assumes that the signal
agrees to within $\sim90\%$ of the template waveform and that the nearby noise
is Gaussian.  We assume all $\theta_i$ are equally likely for signals.

The noise is assumed to be uncorrelated between detectors. The denominator of
the likelihood ratio therefore factors into the product of the distribution of
noise triggers in each detector, $p(\stats_d|\theta_i,\noise)$.  We estimate
these using a two-dimensional kernel density estimation \cite{KDE} constructed from all
of the single-detector triggers not found in coincidence in a single bin.
\begin{figure*}
\includegraphics[width=\columnwidth]{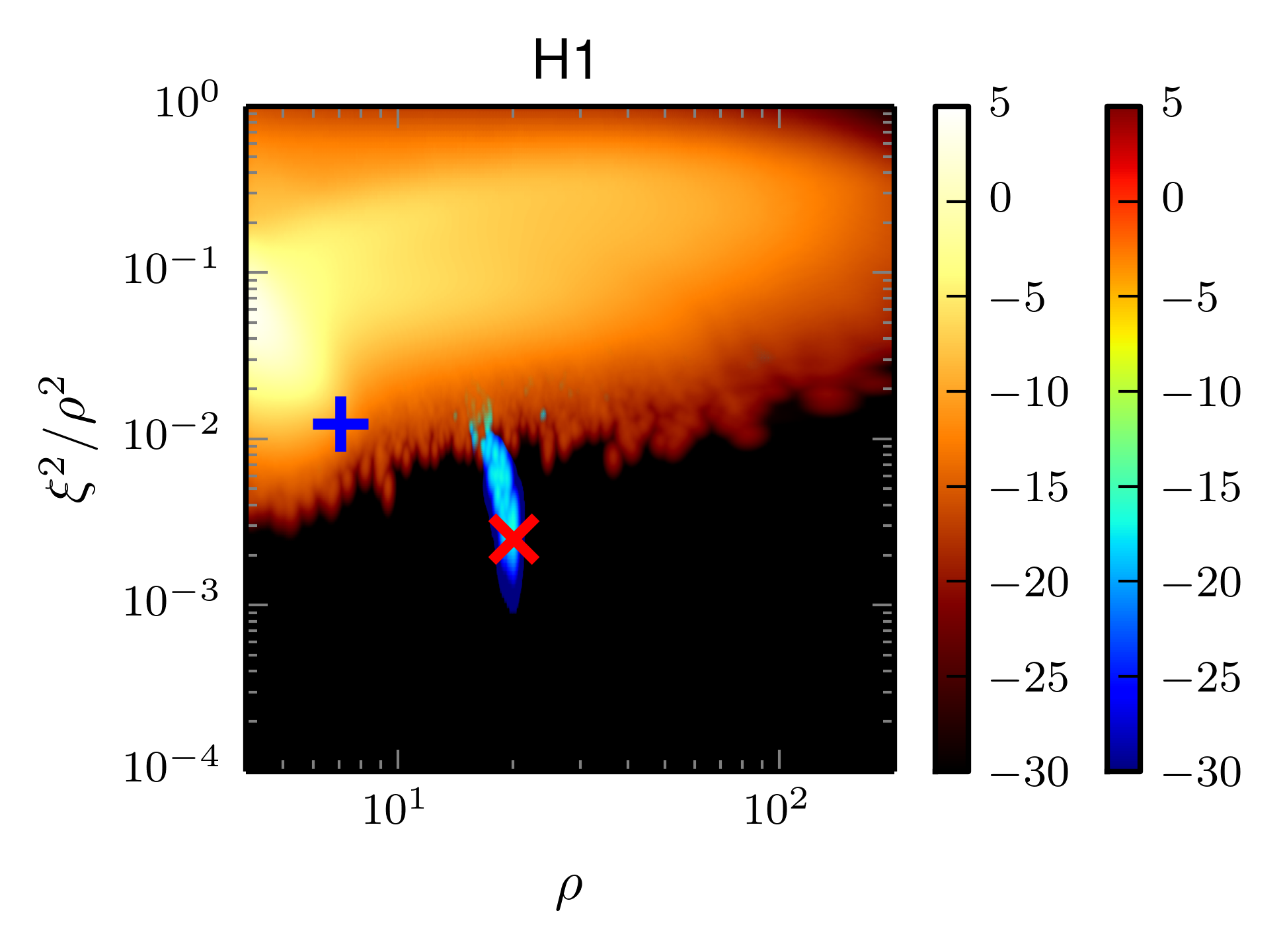}
\includegraphics[width=\columnwidth]{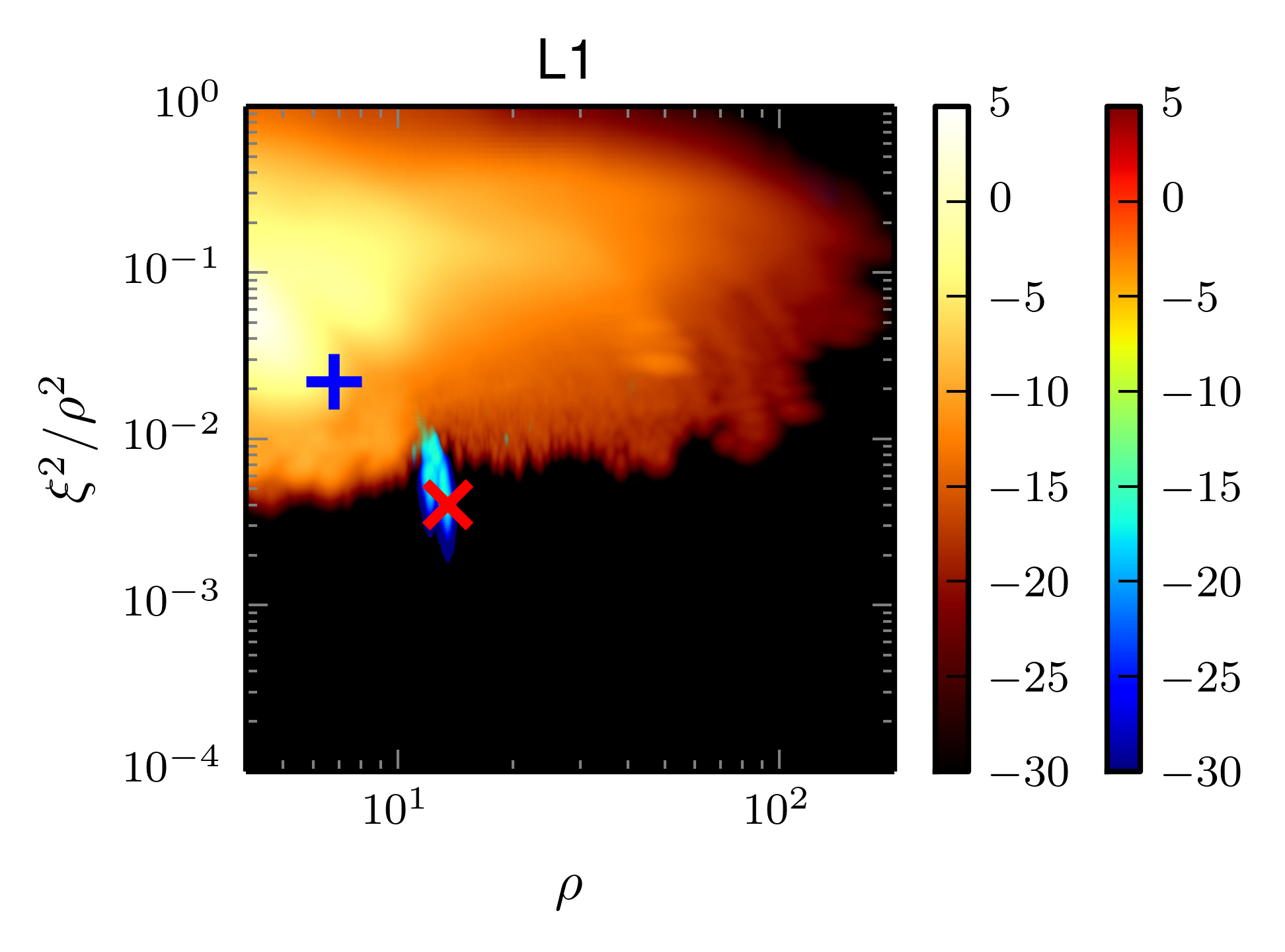}
\caption{ \label{f:gstlalCOMPARISON}
Two projections of parameters in the multi-dimensional likelihood ratio
ranking for \gstlal{} (Left: H1, Right: L1). The relative positions of
\TheEvent{} (red cross) and \SECONDMONDAY{} (blue plus) are indicated in the
$\gstlalXsq/\rho^2$ vs matched-filter SNR plane.  The yellow--black colormap
shows the natural logarithm of the probability density function calculated
using only coincident triggers that are not coincident between the detectors.
This is the background model used in the likelihood ratio calculation. The
red--blue colormap shows the natural logarithm of the probability density
function calculated from both coincident events and triggers that are not
coincident between the detectors. The distribution showing both candidate
events and non-coincident triggers has been masked to only show regions which
are not consistent with the background model. Rather than showing the
$\theta_i$ bins in which \TheEvent{} and \SECONDMONDAY{} were found,
$\theta_i$ has been marginalized over to demonstrate that no background
triggers from any bin had the parameters of \TheEvent{}.}
\end{figure*}

The likelihood ratio \gstlalLR{} provides a ranking of events such that larger
values of \gstlalLR{} are associated with a higher probability of the data
containing a signal. The likelihood ratio itself is not the probability that
an event is a signal, nor does it give the probability that an event was
caused by noise.  Computing the probability that an event is a signal requires
additional prior assumptions. Instead, for each candidate event, we compute
the false alarm probability \fap{}. This is the probability of finding one or
more noise background events with a likelihood-ratio value greater than or
equal to that of the candidate event. Assuming Poisson statistics for the
background, this is given by:
\begin{equation}
\label{eq:fap}
\fap(\gstlalLR) \equiv P(\gstlalLR | T, \noise) = 1 - \exp[-\lambda(\gstlalLR | T, \noise)].
\end{equation}
Instead of using time shifts, the \gstlal{} anlysis estimates the Poisson rate
of background events $\lambda(\gstlalLR{}| T, \noise)$ as:
\begin{equation}
\label{eqn:gstlal_lambda}
\lambda(\gstlalLR{}|T, \noise) = M(T) P(\gstlalLR{}|\noise),
\end{equation}
where $M(T)$ is the number of coincident events found above threshold in the
analysis time $T$, and $P(\gstlalLR{}|\noise)$ is the probability of obtaining
one or more events from noise with a likelihood ratio $\geq \gstlalLR{}$ (the
survival function). We find this by estimating the survival function in each
template bin, then marginalize over the bins; i.e., $P(\gstlalLR{}|\noise) =
\sum_i P(\gstlalLR{}|\theta_i, \noise) P(\theta_i|\noise)$. In a single bin,
the survival function is
\begin{equation}
\label{eqn:gstlal_survivalfunc}
P(\gstlalLR{}|\theta_i,\noise) = 1 - \int_{\intRegion(\gstlalLR{})} p'(\stats_{\Hone}|\theta_i,\noise)p'(\stats_{\Lone}|\theta_i,\noise) \mathrm{d}\stats_{\Hone}\mathrm{d}\stats_{\Lone}.
\end{equation}
Here, $p'(\stats_{d}|\theta_i,\noise)$ are estimates of the distribution of
triggers in each detector including all of the single-detector triggers,
whereas the estimate of $p(\stats_{d}|\theta_i,\noise)$ includes only those
triggers which were not coincident. This is consistent with the assumption
that the false alarm probability is computed assuming all events are noise.

The integration region $\intRegion(\gstlalLR{})$ is the volume of the
four-dimensional space of $\stats_d$ for which the likelihood ratios are less
than $\gstlalLR{}$. We find this by Monte Carlo integration of our estimates
of the single-detector noise distributions $p'(\stats_d|\theta_i,\noise)$.
This is approximately equal to the number of coincidences that can be formed
from the single-detector triggers with likelihood ratios $\geq \gstlalLR{}$
divided by the total number of possible coincidences. We therefore reach a
minimum possible estimate of the survival function, without extrapolation, at
the \gstlalLR{} for which
$p'(\stats_{\Hone}|\theta_i,\noise)p'(\stats_{\Lone}| \theta_i,\noise) \sim
1/N_{\Hone}(\theta_i)N_{\Lone}(\theta_i)$, where $N_d(\theta_i)$ are the total
number of triggers in each detector in the $i$th bin.  

\TheEvent{} was more significant than any other combination of triggers.  For
that reason, we are interested in knowing the minimum false alarm probability
that can be computed by the \gstlal{} analysis.  All of the triggers in a
template bin, regardless of the template from which they came, are used to
construct the single-detector probability density distributions $p'$ within
that bin. The false alarm probability estimated by the \gstlal{} analysis is
the probability that noise triggers occur within a $\pm15\,$ms time window
\emph{and} occur in the same template.  Under the assumption that triggers are
uniformly distributed over the bins, the minimum possible false alarm
probability that can be computed is $M N_{\text{bins}} / (N_{\Hone}
N_{\Lone})$, where $N_{\text{bins}}$ is the number of bins used, $N_{\Hone}$
is the total number of triggers in \Hone{}, and $N_{\Lone}$ is the total
number of triggers in \Lone{}. For the present analysis, $M \sim 1\times10^9$,
$N_{\Hone} \sim N_{\Lone} \sim 1\times10^{11}$, and $N_{\text{bins}}$ is 248,
yielding a minimum value of the false alarm probability of $\sim 10^{-11}$.  

We cannot rule out the possibility that noise produced by the detectors
violates the assumption that it is uniformly distributed among the templates
within a bin.  If we consider a more conservative noise hypothesis that does
not assume that triggers are uniformly distributed within a bin and instead
considers each template as a separate $\theta_i$ bin, we can evaluate the
minimum upper bound on the false alarm probability of \TheEvent{}.  This
assumption would produce a larger minimum false alarm probability value by
approximately the ratio of the number of templates to the present number of
bins.  Under this noise hypothesis, the minimum value of the false alarm
probability would be $\sim 10^{-8}$, which is consistent with the minimum
false alarm probability bound of the \pycbc{} analysis.

Figure~\ref{f:gstlalCOMPARISON} shows $p(\stats_{\Hone}|\noise)$ and
$p(\stats_{\Lone}|\noise)$ in the warm colormap.  The cool colormap includes
triggers that are also found in coincidence, i.e., $p'(\stats_{\Hone}|\noise)$
and $p'(\stats_{\Lone}|\noise)$, which is the probability density function
used to estimate $P(\gstlalLR{}|\noise)$.  It has been masked to only show
regions which are not consistent with $p(\stats_{\Hone}|\noise)$ and
$p(\stats_{\Lone}|\noise)$.  In both cases $\theta_i$ has been marginalized
over in order to show all the data on a single figure.  The positions of the
two loudest events, described in the next section, are shown.  
Figure~\ref{f:gstlalCOMPARISON} shows that \TheEvent{} falls in a region without any
non-coincident triggers from any bin.

\section{Search Results}
\label{s:results}

GW150914 was observed on \OBSEVENTDATEMONTHDAYYEAR\ at \OBSEVENTTIME
~\OBSEVENTTZ\ as the most significant event by both analyses. The individual
detector triggers from GW150914 occurred within the 10\,ms inter-site
propagation time with a combined matched-filter SNR of
\OBSEVENTAPPROXCOMBINEDSNR. Both pipelines report the same matched-filter SNR
for the individual detector triggers in the Hanford detector
($\rho_\textrm{H1} = 20$) and the Livingston detector ($\rho_\textrm{L1} =
13$).  GW150914 was found with the same template in both analyses with
component masses $\CBCEventTemplateMassOne\,\Msun$ and
$\CBCEventTemplateMassTwo\,\Msun$. The effective spin of the best-matching
template is $\chi_{\text{eff}} = (c/G) (\mathbf{S_1}/m_1 + \mathbf{S_2}/m_2)
\cdot (\mathbf{\hat{L}}/M) = 0.2$, where $\mathbf{S_{1,2}}$ are the spins of
the compact objects and $\mathbf{\hat{L}}$ is the direction of the binary's
orbital angular momentum. Due to the discrete nature of the template bank,
follow-up parameter estimation is required to accurately determine the best
fit masses and spins of the binary's components~\cite{Veitch:2014wba,
GW150914-PARAMESTIM}. 

\begin{figure*}[t!]
\centering
\includegraphics[width=\columnwidth]{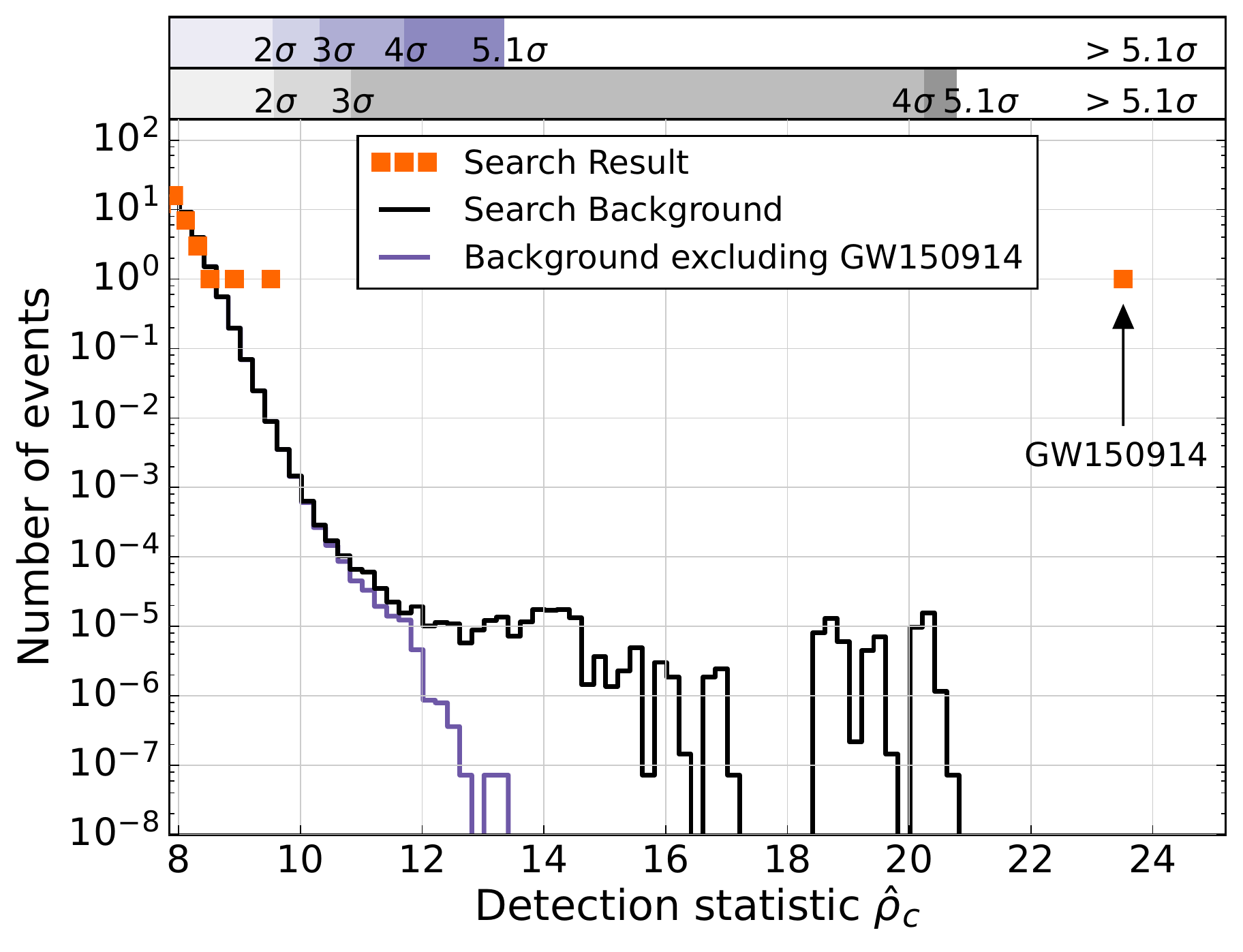}
\includegraphics[width=\columnwidth]{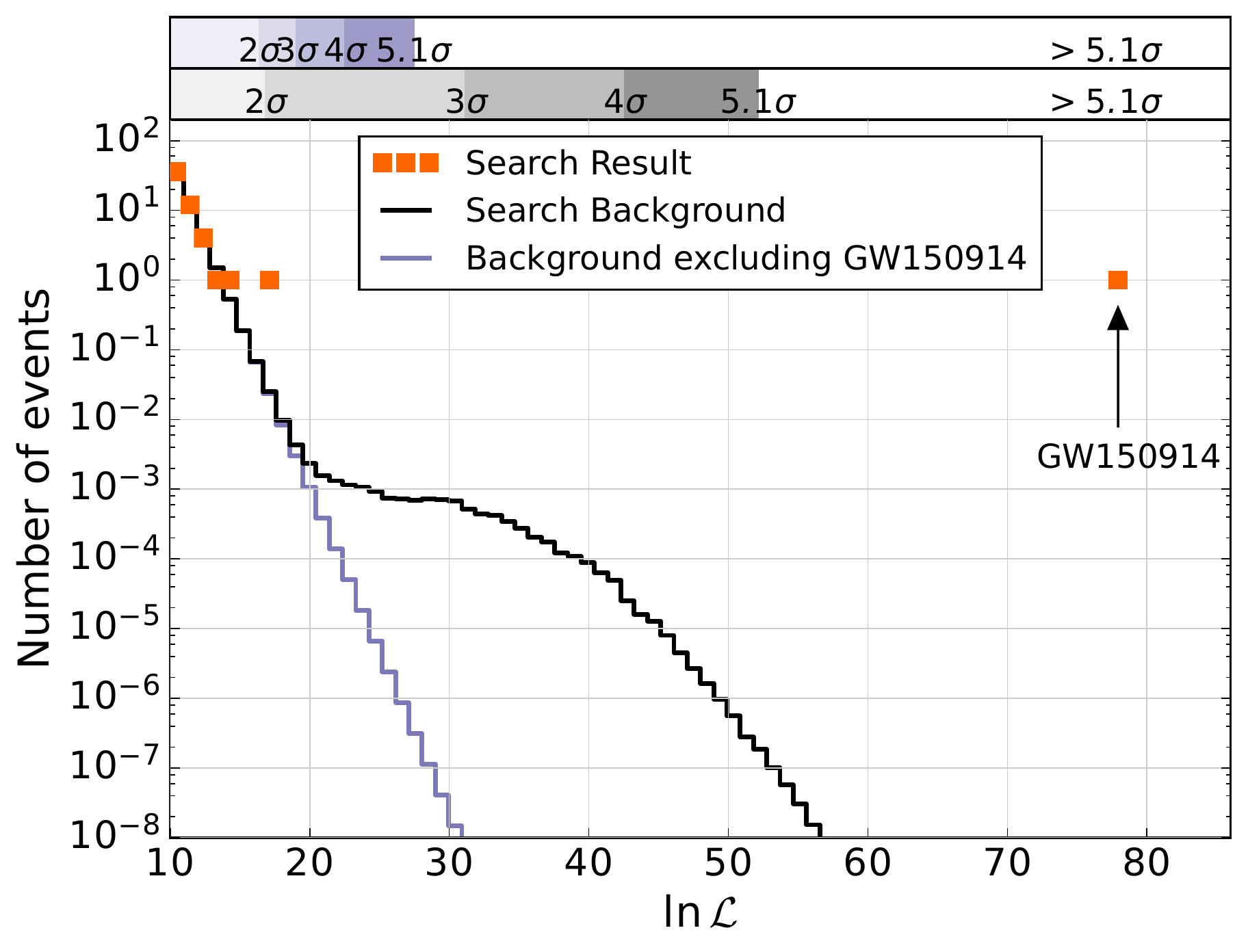}
\caption{Left: Search results from the \pycbc{} analysis. The histogram shows
the number of candidate events (orange) and the number of background events
due to noise in the search class where GW150914 was found (black) as a
function of the search detection-statistic and with a bin width of $\Delta
\hat{\rho}_c = 0.2$. The significance of GW150914 is greater than
{\CBCEVENTSIGMA} $\sigma$.  The scales immediately above the histogram give
the significance of an event measured against the noise backgrounds in units
of Gaussian standard deviations as a function of the detection-statistic.  The
black background histogram shows the result of the time-shift method to
estimate the noise background in the observation period. The tail in the
black-line background of the binary coalescence search is due to random
coincidences of GW150914 in one detector with noise in the other detector. The
significance of GW150914 is measured against the upper gray scale.  The purple
background histogram is the background excluding coincidences involving
GW150914 and it is the background to be used to assess the significance of the
second loudest event; the significance of this event is measured against the
upper purple scale.  Right: Search results from the  \gstlal{} analysis. The
histogram shows the observed candidate events (orange) as a function of the
detection statistic $\ln \mathcal{L}$. The black line indicates the expected
background from noise where candidate events have been included in the noise
background probability density function. The purple line indicates the
expected background from noise where candidate events have not been included in
the noise background probability density function. The
independently-implemented search methods and different background estimation
method confirm the discovery of GW150914.}
\label{fig:both_results}
\end{figure*}

The frequency at peak amplitude of the best-matching template is
$f_{\mathrm{peak}} = \CBCEventPeakFrequency\,$Hz, placing it in
noise-background class (iii) of the \pycbc{} analysis.
Figure~\ref{fig:both_results} (left) shows the result of the \pycbc{} analysis
for this search class.  In the time-shift analysis used to create the noise
background estimate for the \pycbc{} analysis, a signal may contribute events
to the background through random coincidences of the signal in one
detector with noise in the other detector~\cite{Capano:2016uif}.  This can be
seen in the background histogram shown by the black line. The tail is due
to coincidence between the single-detector triggers from GW150914 and noise in
the other detector.  If a loud signal is in fact present, these random
time-shifted coincidences contribute to an overestimate of the noise
background and a more conservative assessment of the significance of an event.
Figure~\ref{fig:both_results} (left) shows that GW150914 has a re-weighted SNR
$\rwRhoC = \PycbcEventNewSNR$, greater than all background events in its
class. This value is also greater than all background in the other two
classes. As a result, we can only place an upper bound on the false alarm
rate, as described in Sec.~\ref{s:pycbc}. This bound is equal to the number of
classes divided by the background time $T_b$. With $3$ classes and $T_b =
\PYCBCBCKLIVETIME$ years, we find the false alarm rate of GW150914 to be less
than $\CBCEVENTFAR\,\text{yr}^{-1}$. With an observing time of $\OBSHOURS$,
the false alarm probability is $\fap~\CBCEVENTFAPBOUND$.  Converting this
false alarm probability to single-sided Gaussian standard deviations according
to $-\sqrt{2}\ \mathrm{erf}^{-1}\left[1 - 2(1-\fap)\right]$, where
$\mathrm{erf}^{-1}$ is the inverse error function, the \pycbc{} analysis
measures the significance of GW150914 as greater than \CBCEVENTSIGMA
$\,\sigma$. 

\newcolumntype{M}[1]{>{\centering\arraybackslash}m{#1}}
\begin{table*}[htb]
  \begin{ruledtabular}
  \begin{tabular}{M{1.6cm} M{1.8cm} M{1.5cm} M{1.5cm} M{1.5cm} M{1.5cm} M{1.5cm} M{1.8cm} M{1.5cm}}
    Event & Time (UTC) & FAR (yr$^{-1}$) & \fap{} & $\mathcal{M}$ $(\Msun)$ & $m_1$ $(\Msun)$ & $m_2$ $(\Msun)$ & $\chi_{\mathrm{eff}}$ & $D_L$ (Mpc) \\
    \hline
    \TheEvent{} & \CBCEventUTCTimeShort & $< \CBCEVENTFAR$ & $\CBCEVENTFAPBOUND$\newline$(> \CBCEVENTSIGMA\,\sigma)$ & \MCSCOMPACT & \MONESCOMPACT & \MTWOSCOMPACT &  \CHIEFFCOMPACT & \DISTANCECOMPACT \\
    \SECONDMONDAY{} & \CBCSecondEventUTCTimeShort & \CBCSECONDEVENTFAR & $\CBCSECONDEVENTFAP$\newline$(\PyCBCSecondEventSigma\,\sigma)$ &  \MCSCOMPACTSecondMonday & \MONESCOMPACTSecondMonday  & \MTWOSCOMPACTSecondMonday & \CHIEFFCOMPACTSecondMonday & \DISTANCECOMPACTSecondMonday \\
  \end{tabular}
  \end{ruledtabular}
\caption{
\label{tab:results}
Parameters of the two most significant events. The false alarm rate (FAR) and
false alarm probability (\fap{}) given here were determined by the \pycbc{}
pipeline; the \gstlal{} results are consistent with this. The source-frame
chirp mass $\mathcal{M}$, component masses $m_{1,2}$, effective spin
$\chi_{\mathrm{eff}}$, and luminosity distance $D_L$ are determined using a
parameter estimation method that assumes the presence of a coherent compact
binary coalescence signal starting at 20\,Hz in the
data~\cite{Veitch:2014wba}. The results are computed by averaging the
posteriors for two model waveforms.  Quoted uncertainties are $90\%$ credible
intervals that include statistical errors and systematic errors from
averaging the results of different waveform models.  Further parameter
estimates of \TheEvent{} are presented in Ref.~\cite{GW150914-PARAMESTIM}.
}
\end{table*}

\begin{figure*}[t]
\includegraphics[width=\columnwidth]{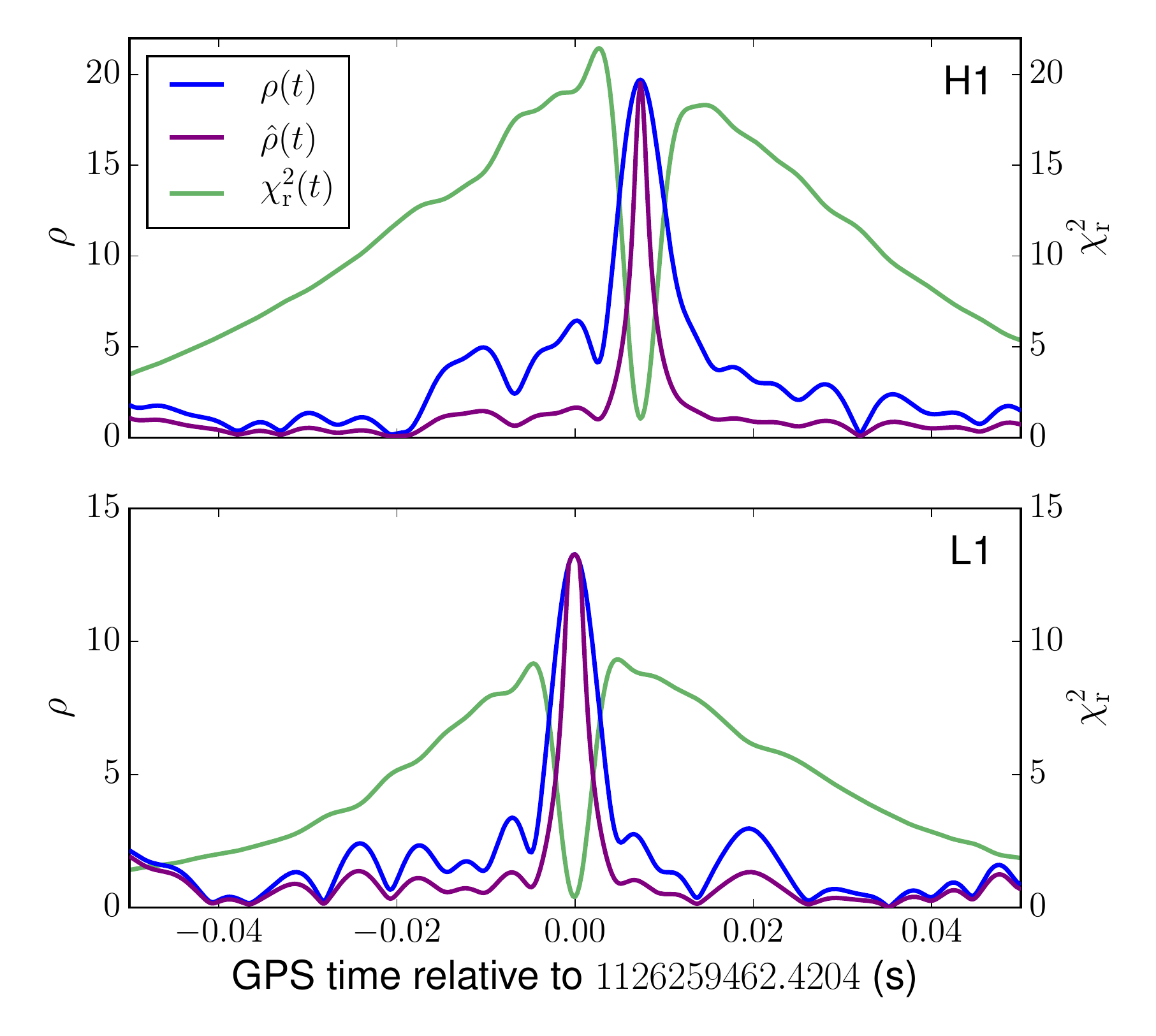}
\includegraphics[width=0.931\columnwidth]{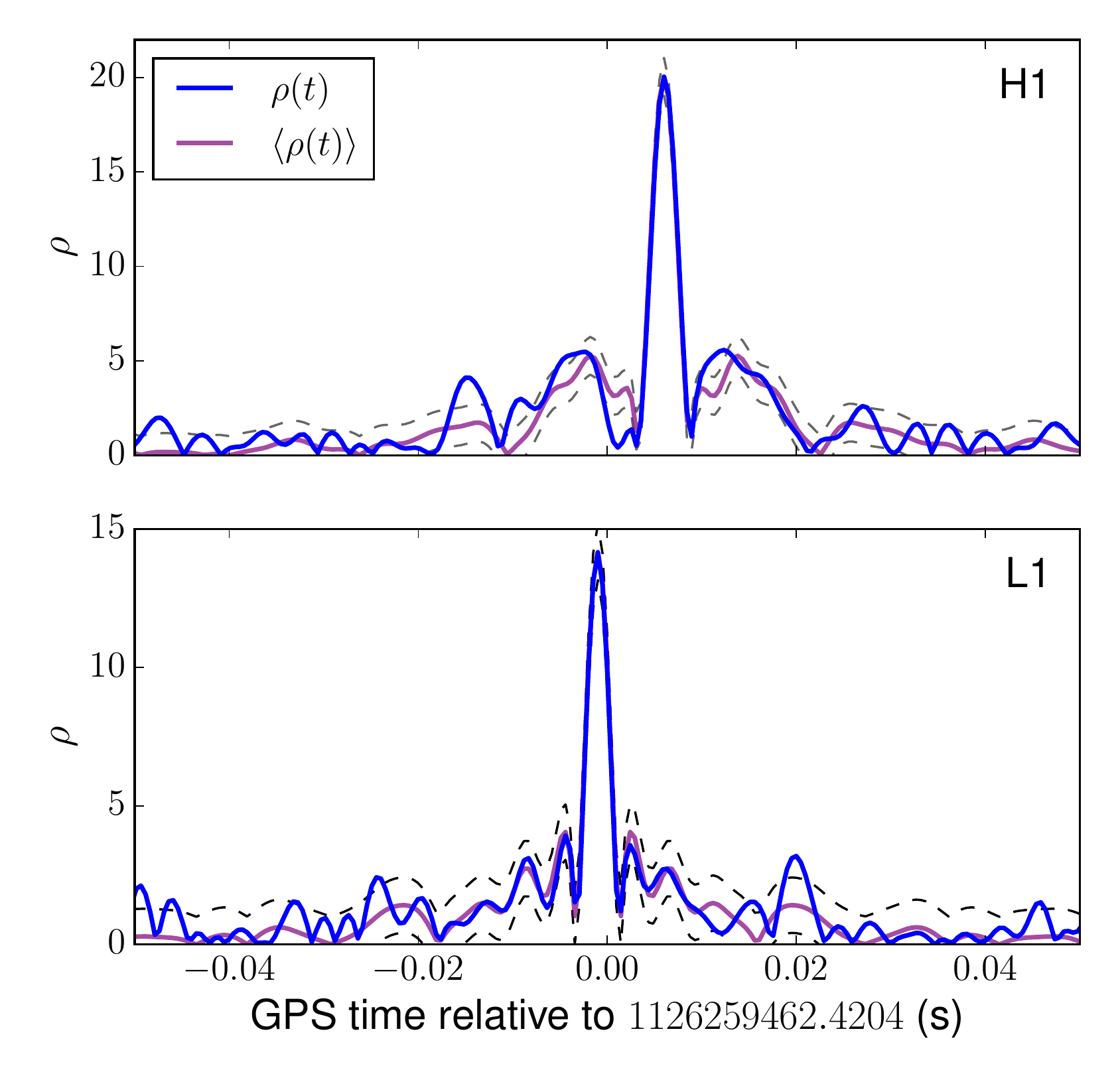} 
\caption{\label{fig:pycbc_gstlal_snr_chisq_vs_time}
Left: PyCBC matched-filter SNR (blue), re-weighted SNR (purple) and $\chi^2$
(green) versus time of the best-matching template at the time of \TheEvent.
The top plot shows the Hanford detector; bottom, Livingston.  Right: Observed
matched-filter SNR (blue) and expected matched-filter SNR (purple) versus time
for the best-matching template at the time of \TheEvent, as reported by the
\gstlal{} analysis. The expected matched-filter SNR is based on the
autocorrelation of the best-matching template.  The dashed black lines
indicate $1\sigma$ deviations expected in Gaussian noise.
}
\end{figure*}

The \gstlal{} analysis reported a detection-statistic value for GW150914 of
$\ln{\gstlalLR{}} = 78$, as shown in the right panel of
Fig.~\ref{fig:both_results}. The \gstlal{} analysis estimates the false alarm
probability assuming that noise triggers are equally likely to occur in any of
the templates within a background bin.  However, as stated in Sec.
\ref{s:gstlal}, if the distribution of noise triggers is not uniform across
templates, particularly in the part of the bank where \TheEvent{} is observed,
the minimum false alarm probability would be higher.  For this reason we quote
the more conservative \pycbc{} bound on the false alarm probability of
\TheEvent{} here and in Ref.~\cite{GW150914-DETECTION}.  However, proceeding
under the assumption that the noise triggers are equally likely to occur in
any of the templates within a background bin, the \gstlal{} analysis estimates
the false alarm probability of \TheEvent{} to be \gstLALEventFAP.  The
significance of GW150914 measured by \gstlal{} is consistent with the bound
placed by the \pycbc{} analysis and provides additional confidence in the
discovery of the signal.

The difference in time of arrival between the Livingston and Hanford detectors
from the individual triggers in the \pycbc{} analysis is
$\CBCEventTimeDiff\,$ms, consistent with the time delay of
\TIMEDELAYCOMPACT\,ms recovered by parameter
estimation~\cite{GW150914-PARAMESTIM}.
Figure~\ref{fig:pycbc_gstlal_snr_chisq_vs_time} (left) shows the
matched-filter SNR $\rho$, the $\chi^2$-statistic, and the re-weighted SNR
$\hat\rho$ for the best-matching template over a period of $\pm5$\,ms around
the time of GW150914 (we take the \pycbc{} trigger time in L1 as a reference).
The matched-filter SNR peaks in both detectors at the time of the event and
the value of the reduced chi-squared statistic is $\chi^2_\mathrm{H1} = 1$ and
$\chi^2_\mathrm{L1} = 0.7$ at the time of the event, indicating an excellent
match between the template and the data. The re-weighted SNR of the individual
detector triggers of $\hat{\rho}_\mathrm{H1} = 19.5$ and
$\hat{\rho}_\mathrm{L1} = 13.3$ are larger than that of any other
single-detector triggers in the analysis; therefore the significance
measurement of \CBCEVENTSIGMA $\,\sigma$ set using the $0.1$~s time shifts is
a conservative bound on the false alarm probability of \TheEvent.

Figure~\ref{fig:pycbc_gstlal_snr_chisq_vs_time} (right) shows $\pm5$\,ms of the
\gstlal{} matched-filter SNR time series from each detector around the event
time together with the predicted SNR time series computed from the
autocorrelation function of the best-fit template.  The difference between the
autocorrelation and the observed matched-filter SNR is used to perform the
\gstlal{} waveform-consistency test.  The autocorrelation matches the observed
matched-filter SNR extremely well, with consistency test values of
$\xi_\mathrm{H1} = 1$ and $\xi_\mathrm{L1} = 0.7$.  No other triggers with
comparable matched-filter SNR had such low values of the signal-consistency
test during the entire observation period.

Both analyses have shown that the probability that GW150914 was formed by
random coincidence of detector noise is extremely small. We therefore conclude
that GW150914 is a gravitational-wave signal. To measure the signal
parameters, we use parameter estimation methods that assume the presence of a
coherent coalescing binary signal in the data from both
detectors~\cite{Veitch:2014wba,GW150914-PARAMESTIM}. Two waveform models were
used which included inspiral, merger and ring-down portions of the signal: one
which includes spin components aligned with orbital angular
momentum~\cite{Pan:2009wj,Puerrer:2014fza} and one which includes the dominant
modulation of the signal due to orbital precession caused by mis-aligned
spins~\cite{Hannam:2013oca,Khan:2015jqa}.  The parameter estimates are
described by a continuous probability density function over the source
parameters. We conclude that \TheEvent{} is a nearly equal mass black-hole
binary system of source-frame masses {\MONESCOMPACT~\Msun} and
{\MTWOSCOMPACT~\Msun} (median and $90\%$ credible range). The spin magnitude
of the primary black hole is constrained to be less than ${\SPINONELIMIT}$
with $90\%$ probability.  The most stringent constraint on the spins of the
two black holes is on the effective spin parameter $\chi_\mathrm{eff} =
\CHIEFFCOMPACT$.  The parameters of the best-fit template are consistent with
these values, given the discrete nature of the template bank.  We estimate
\TheEvent{} to be at a luminosity distance of
{\DISTANCECOMPACT~$\mathrm{Mpc}$}, which corresponds to a redshift
{\REDSHIFTCOMPACT}.  Full details of the source parameters for GW150914 are
given in Ref.~\cite{GW150914-PARAMESTIM} and summarized in
Table~\ref{tab:results}.

When an event is confidently identified as a real gravitational wave signal,
as for GW150914, the background used to determine the significance of other
events is re-estimated without the contribution of this event. This is the
background distribution shown as purple lines in Fig.~\ref{fig:both_results}.
Both analyses reported a candidate event on \SecondTime{} as the
second-loudest event in the observation period, which we refer to as
\SECONDMONDAY. This candidate event has a combined matched-filter SNR of
\PyCBCSecondEventRhoC.  The \pycbc{} analysis reported a false alarm rate of 1
per \CBCSECONDEVENTIFAR~years and a corresponding false alarm probability of
\CBCSECONDEVENTFAP{} for this event.  The \gstlal{} analysis reported a false
alarm rate of 1 per \gstLALSecondEventInverseFAR{}~years and a false alarm
probability of \gstLALSecondEventFAP.  These results are consistent with
expectations for candidate events with low matched-filter SNR, since \pycbc{}
and \gstlal{} use different ranking statistics and background estimation
methods.  Detector characterization studies have not identified an
instrumental or environmental artifact as causing this candidate event
\cite{GW150914-DETCHAR}, however its false alarm probability is not
sufficiently low to confidently claim the event as a signal.  It is
significant enough to warrant follow-up, however.  The results of signal
parameter estimation, shown in Table~\ref{tab:results}, indicate that if
\SECONDMONDAY{} is of astrophysical origin, then the source would be a
stellar-mass binary black hole system with source-frame component masses
$\MONESCOMPACTSecondMonday\,\Msun$ and $\MTWOSCOMPACTSecondMonday\,\Msun$. The
effective spin would be $\chi_\mathrm{eff} = \CHIEFFCOMPACTSecondMonday$ and
the distance \DISTANCECOMPACTSecondMonday~Mpc.

\section{Conclusion}
\label{s:conclusion}

The LIGO detectors have observed gravitational waves from the merger of two
stellar-mass black holes. The binary coalescence search detects GW150914 with
a significance greater than $\CBCEVENTSIGMA$$\sigma$ during the observations
reported. This result is confirmed by two independent matched filter analyses,
providing confidence in the discovery. Detailed parameter estimation for
GW150914 is reported in Ref.~\cite{GW150914-PARAMESTIM}, the implications for
the rate of binary black hole coalescences in Ref.~\cite{GW150914-RATES}, and
tests for consistency of the signal with general relativity in
Ref.~\cite{GW150914-TESTOFGR}.  Ref.~\cite{GW150914-ASTRO} discusses the
astrophysical implications of this discovery.  Full results of the compact
binary search in Advanced LIGO's first observing run will be reported in a
future publication.

\acknowledgments

The authors gratefully acknowledge the support of the United States National
Science Foundation (NSF) for the construction and operation of the LIGO
Laboratory and Advanced LIGO as well as the Science and Technology Facilities
Council (STFC) of the United Kingdom, the Max-Planck-Society (MPS), and the
State of Niedersachsen/Germany for support of the construction of Advanced
LIGO and construction and operation of the GEO\,600 detector.  Additional
support for Advanced LIGO was provided by the Australian Research Council.
The authors gratefully acknowledge the Italian Istituto Nazionale di Fisica
Nucleare (INFN),  the French Centre National de la Recherche Scientifique
(CNRS) and the Foundation for Fundamental Research on Matter supported by the
Netherlands Organisation for Scientific Research, for the construction and
operation of the Virgo detector and the creation and support  of the EGO
consortium.  The authors also gratefully acknowledge research support from
these agencies as well as by the Council of Scientific and Industrial Research
of India, Department of Science and Technology, India, Science \& Engineering
Research Board (SERB), India, Ministry of Human Resource Development, India,
the Spanish Ministerio de Econom\'ia y Competitividad, the Conselleria
d'Economia i Competitivitat and Conselleria d'Educaci\'o, Cultura i
Universitats of the Govern de les Illes Balears, the National Science Centre
of Poland, the European Commission, the Royal Society, the Scottish Funding
Council, the Scottish Universities Physics Alliance, the Hungarian Scientific
Research Fund (OTKA), the Lyon Institute of Origins (LIO), the National
Research Foundation of Korea, Industry Canada and the Province of Ontario
through the Ministry of Economic Development and Innovation, the National
Science and Engineering Research Council Canada, Canadian Institute for
Advanced Research, the Brazilian Ministry of Science, Technology, and
Innovation, Russian Foundation for Basic Research, the Leverhulme Trust, the
Research Corporation, Ministry of Science and Technology (MOST), Taiwan and
the Kavli Foundation.  The authors gratefully acknowledge the support of the
NSF, STFC, MPS, INFN, CNRS and the State of Niedersachsen/Germany for
provision of computational resources.

\appendix

\section{Detector Calibration}
\label{s:calibration}

The LIGO detectors do not directly record the strain signal, rather they sense
power fluctuations in the light at the interferometer's readout
port~\cite{TheLIGOScientific:2014jea}.  This error signal is used to generate
a feedback signal to the detector's differential arm length to maintain
destructive interference of the light moving towards the readout
port~\cite{GW150914-DETECTORS}. The presence of this feedback signal
suppresses the length change from external sources; a combination of the error
and control signals is used to estimate the detector strain.  The strain is
calibrated by measuring the detector's response to test mass motion induced by
photon pressure from a modulated calibration laser beam.  Changes in the
detector's thermal and alignment state cause small, time-dependent systematic
errors in the calibration. For more details see
Ref.~\cite{GW150914-CALIBRATION}.

Errors in the calibrated strain data lead to mismatches between waveform
templates and the gravitational-wave signal. This mismatch has been shown to
decrease the expected SNR $\langle\rho\rangle$, but only has a weak, quadratic
dependence on calibration errors~\cite{Allen:1996,Brown:2004}. However, the
quantity used for detection is the re-weighted SNR $\hat\rho(\rho, \chi_r^2)$
for each detector. In this appendix, we analyze the impact of calibration
errors on $\hat\rho$ for signals similar to GW150914, and we find that the
expected re-weighted SNR $\langle\hat\rho\rangle$ also shows only a weak
dependence on calibration errors.  

In the frequency domain, the process of calibration reconstructs the
gravitational-wave strain $h(f) = \Delta L(f)/L$ from the differential arm
length error signal $d_{\rm err}(f)$, which is the filtered output of the
photodiode. The function that relates the two quantities is the response
function $R(f)$
\begin{equation}
\Delta L(f) = R(f) d_{\rm err}(f),
\end{equation}
This response function is constructed from the sensing transfer function
$C(f)$ that describes the frequency response of the detector to changes in the
arm lengths as well as the actuation transfer function $A(f)$ that describes
the motion of the test mass when driven by the control signal to maintain
destructive interference in the interferometer~\cite{GW150914-CALIBRATION}.

The initial sensing and actuation transfer functions, measured before the
start of the observing run, are defined by $C_0(f)$ and $A_0(f)$ respectively.
However, over the course of an observing run, the frequency dependence of
these transfer functions slowly drift. The drift in the sensing function is
parameterized by the real correction factor $\kappa_C$ and the cavity pole
frequency $f_C$, while the drift in the actuation function is parameterized by
the complex correction factor for the actuation of the test mass $\kappa_T$ as
well as by the complex correction factor for the penultimate and
upper-intermediate masses of the test-mass suspension system
$\kappa_{PU}$~\cite{GW150914-CALIBRATION}.  This results in six real
time-dependent parameters: \{$\Re\kappa_T$, $\Im\kappa_T$, $\Re\kappa_{PU}$,
$\Im\kappa_{PU}$, $\kappa_C$, $f_C$\}, with nominal values $\kappa_C = 1$,
$\kappa_{T} = 1$, and $\kappa_{PU} = 1$ for the correction factors, as well as
the cavity pole frequencies $f_C = \FCHanford$~Hz for LHO and $f_C =
\FCLivingston$~Hz for LLO. The drift in these parameters is monitored by
actuating the test masses at specific frequencies called calibration lines
using the photon calibrator and electrostatic drive, and these parameters are
found to deviate from their nominal values by no more than
\paramvary\%~\cite{GW150914-CALIBRATION}.

The analysis for the discovery of GW150914 uses calibrated strain data that
does not correct for these time varying parameters. We can evaluate the impact
on $\hat\rho$ of not including these parameters by adding simulating signals
to the strain data before filtering and adjusting these data with artificial
values of these parameters. We performed software injections at 16 different
times on September 14 and 15 using the best-match template for GW150914 given
by the template waveform with parameters $m_1 = \mo M_\odot$, $m_2 = \mt
M_\odot$, $\chi_1 = \chio$, and $\chi_2 = \chit$. We then vary each of the six
time-dependent parameters and calculate $\hat\rho$ with PyCBC for the
re-calibrated strain.

\begin{figure}[tb]
\centering
\includegraphics[width=0.9\linewidth]{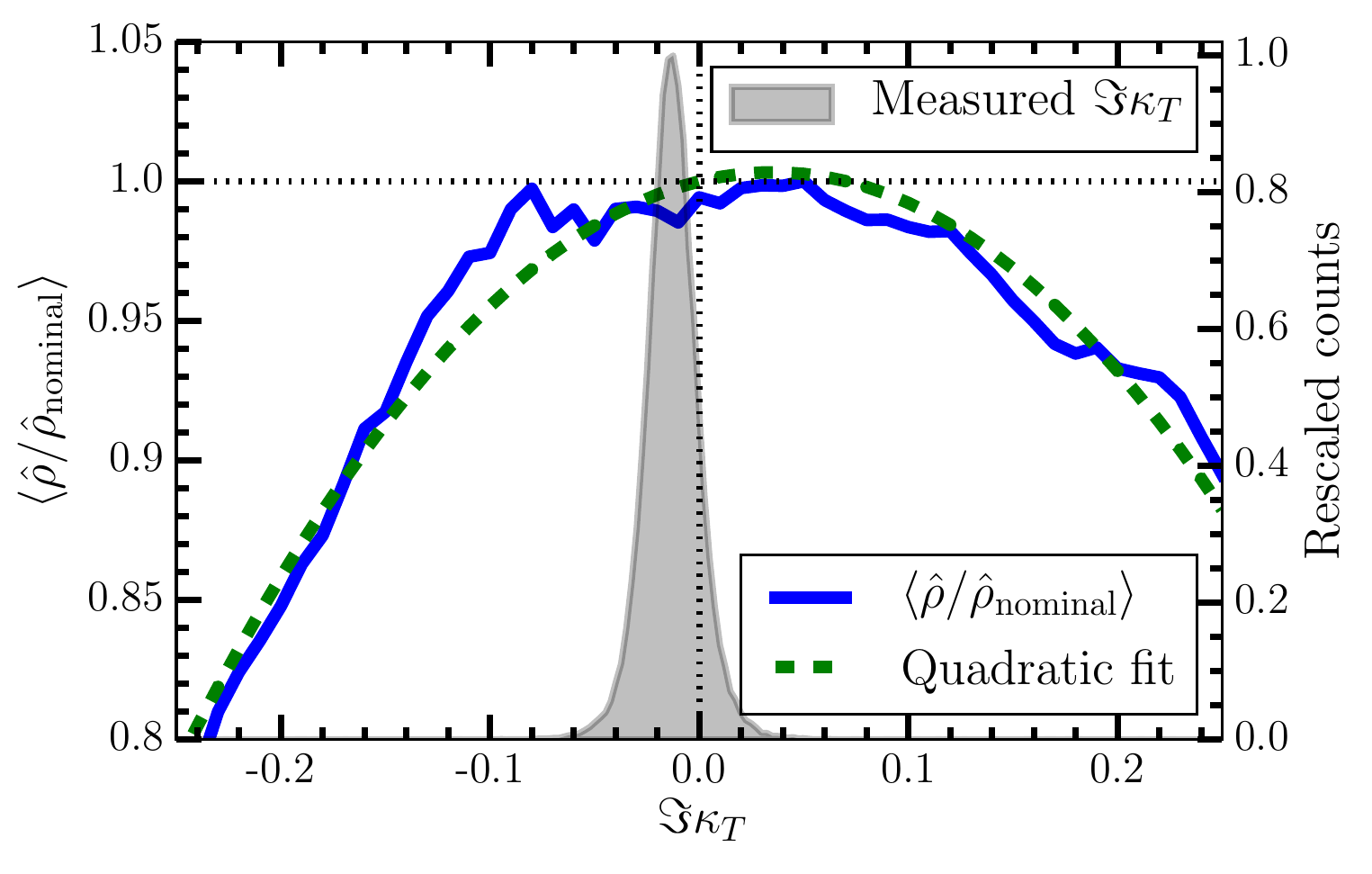}
\caption{Variation in $\hat\rho$ when the time-dependent parameter
$\Im\kappa_T$ is adjusted. The solid blue curve represents $\hat\rho$ averaged
over 16 software injections, with waveform parameters identical to the
best-fit template for GW150914. $\hat\rho$ decreases as $\Im\kappa_{T}$
deviates from its nominal value. The green dashed curve is a quadratic fit,
and represents the approximate behavior of $\langle\hat\rho\rangle$ if we had
used a much larger number of software injections. The grey histogram
represents the measured values of $\Im\kappa_{T}$ for times used in the
analysis on September 14 and 15.} \label{fig:snr_loss} 
\end{figure}

As an example, Fig.~\ref{fig:snr_loss} shows the loss in $\hat\rho$ for LHO as
the parameter $\Im\kappa_{T}$ is artificially adjusted from its nominal value
of zero. Here, $\hat\rho$ is rescaled with respect to its value $\hat\rho_{\rm
nominal}$ at $\Im\kappa_{T}=0$, then averaged over the 16 software injections
to estimate $\langle\hat\rho/\hat\rho_{\rm nominal}\rangle$, the expected
fractional loss in $\hat\rho$. For extreme values of $\Im\kappa_{T}$, the loss
in $\hat\rho$ can be as much as $\sim 20$\%. However, as shown by the
histogram, the measured value of $\Im\kappa_{T}$ rarely deviates by more than
$\imkappaTvary$ from its nominal value, leading to a loss in $\hat\rho$ of no
more than \imkappaTsnrlossH\%. The other five calibration parameters have a
slightly smaller impact on $\hat\rho$. Similar results hold for LLO, except
for variations in $\Im\kappa_{T}$, which can lead to a loss in $\hat\rho$ of
no more than \imkappaTsnrlossL\%.

In addition to the dependence of $\langle\hat\rho\rangle$ on calibration
errors presented in Fig.~\ref{fig:snr_loss}, individual realizations of
$\hat\rho$ show an additional variation of approximately $\pm 2\%$ due to the
power spectral density of the detector noise that is estimated from the strain
data. For example, calibration errors affect the estimated noise power
spectral density, and as a result, shift the bin boundaries used to calculate
the $\chi_r^2$ statistic. This subtle shift in the bin boundaries sometimes
leads to a deviation in the measured value of $\hat\rho$ of about $\pm 2\%$
compared to its value if the bin boundaries had been fixed. The estimate of
the noise power spectral density is also affected by the choice of start and
end times for the \seglen~s segments used to estimate the noise power spectral
density, and this also affects $\hat\rho$ for GW150914 by approximately $\pm
2\%$. Overall, since the measured combined re-weighted SNR $\hat\rho_c$ is
significantly above the detection threshold, neglecting the time-variation of
the calibration does not affect the result of this search.

\section{Analysis of Simulated Signals}
\label{s:validation}

\begin{figure*}[htb]
\includegraphics{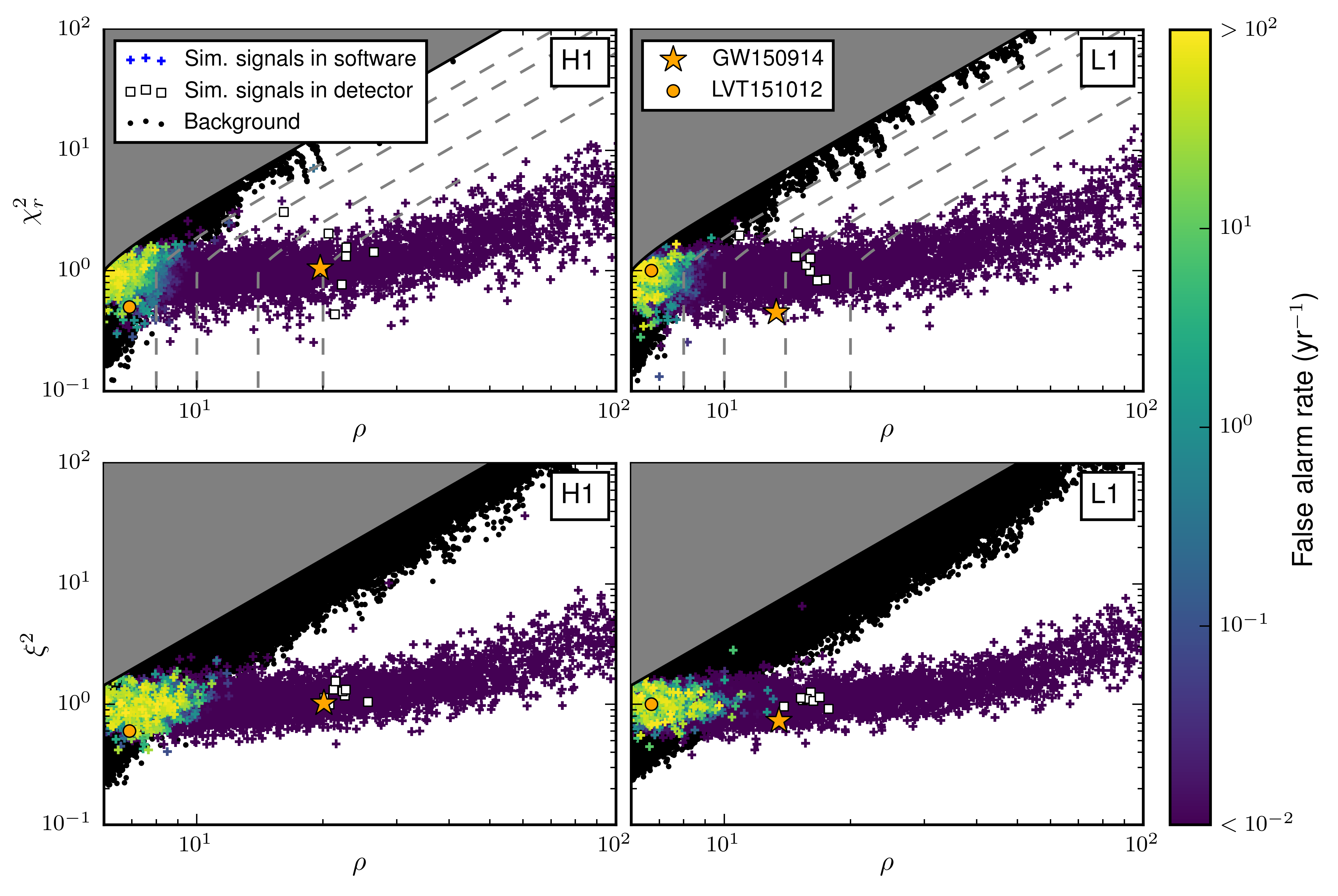}
\caption{\label{fig:snr_chisq} \pycbc{} $\chi_r^2$ (top row) and \gstlal{}
$\xi^2$ (bottom row) versus SNR in each detector. Triggers associated with a
set of simulated binary black hole signals that are added in software are shown, colored by
the false alarm rate that they were recovered with (crosses). Also shown are
triggers associated with simulated signals that were added to the detectors. We
see a clear separation between these simulated signals and background noise
triggers (black dots; for plotting purposes, a threshold was applied to the
background, indicated by the gray region). Lines of constant re-weighted SNR
(gray dashed lines) are shown in the \pycbc{} plot; plotted are $\hat{\rho} =
\{8, 10, 14, 20\}$. 
}
\end{figure*}

Simulated signals are added to the detector data to validate the performance
of our searches. These simulations can be added either in software, by adding
a waveform to the input strain data, or by moving the detector's test masses
in a way that simulates a gravitational-wave signal.  Physically actuating on
the detector's test masses provides a full end-to-end validation of our ability
to detect signals at the expense of corrupting the data during the time of the
simulated signal. Adding simulations in software allows us to repeat the
analysis on the same data set many times, accumulating large statistics and
testing search sensitivity across a large parameter space. Signals simulated
in software have been used to constrain the coalescence rate of binary black
hole systems in Ref.~\cite{GW150914-RATES}.

To validate the search, we generate a population of binary black holes with
component masses between $2$ and $98\,\Msun$ and the full range of available
spins using the template waveform.  Signals are randomly distrubted in sky
location, orientation, distance, and time, then added coherently to each
detector's strain data prior to filtering.  The \pycbc{} and \gstlal{}
analyses report the matched-filter SNR, and the $\chi^2_r$ and \gstlalXsq{}
statistics, respectively, for these simulated signals.  In addition, we
simulate eight  signals in the detectors to test the recovery of GW150914. The
signals were generated using the aligned-spin waveforms used in the search.
The parameters were drawn from the posterior distribution of early GW150914
parameter estimation results. The sky position of the signals were chosen to
give similar amplitudes as GW150914 in the H1 and L1
detectors~\cite{GW150914-PARAMESTIM}.  The signals are added to both detectors
with the correct relative amplitude, phase, and time offsets to simulate a
gravitational-wave signal from an astrophysical source.

Figure~\ref{fig:snr_chisq} shows the $\chi^2_r$ and \gstlalXsq{} versus
matched-filter SNR in each detector for a set of software-simulated binary
black hole signals recovered by the \pycbc{} (top) and \gstlal{} (bottom)
analyses. Also shown are the eight simulated signals that were physically
added to the detector. The parameters of \TheEvent{} are shown with a star. We
see a clear separation between signal and noise background in the region of
GW150914 for both the software and physical (hardware) simulations.  Simulated
gravitational waves with similar parameters and distances as \TheEvent{} are
found with high significance by both analyses, validating the ability of the
analyses described here to detect sources similar to \TheEvent{}.

\end{document}